\documentclass[11pt]{article}
\usepackage{latexsym}

\usepackage{amssymb}
\usepackage{amsmath}

 \usepackage{epsfig}
 \usepackage{graphicx}
\usepackage{hyperref}

\newcommand{\be}{\begin{equation}}
\newcommand{\ee}{\end{equation}}
\newcommand{\ba}{\begin{eqnarray}}
\newcommand{\ea}{\end{eqnarray}}
\newcommand{\baa}{\begin{array}}
\newcommand{\eaa}{\end{array}}
\newcommand{\bi}{\begin{itemize}}
\newcommand{\ei}{\end{itemize}}
\newcommand{\edoc}{\end{document}}

\newcommand{\nn}{\nonumber \\}
\newcommand{\nr}[1]{(\ref{#1})}
\newcommand{\la}[1]{\label{#1}}

\newcommand{\rmi}[1]{{\mbox{\scriptsize #1}}}
\newcommand{\fr}[2]{{\frac{#1}{#2}\,}}
\newcommand{\fra}[2]{{\textstyle{\frac{#1}{#2}\,}}}  

\newcommand{\gev}{{\rm GeV}}

\newcommand{\ph}{\hat p_h}

 \def\l{\lambda}

 \def\e{\epsilon}
 
 \def\p{\partial}
 
 \def\sp{\;\;\;,\;\;\;}
\newcommand{\f}{{\kappa}} 

\def\CO{{\cal O}}
\def\gsim{\gtrsim}
\def\lsim{\lesssim}
\DeclareGraphicsExtensions{.png}

\begin{document}

\begin{titlepage}
\begin{flushright}
HIP-2015-05/TH\\
CCTP-2015-04\\
CCQCN-2015-62
\end{flushright}
\begin{centering}
\vfill

{\Large{\bf Quantum and stringy corrections to the  equation of state of holographic 
QCD matter and the nature of the chiral transition} }

\vspace{0.8cm}

T. Alho$^{\rm a}$,
M. J\"arvinen$^{\rm b}$,
K. Kajantie$^{\rm c}$,        
E. Kiritsis$^{\rm d,e}$,
K. Tuominen$^{\rm c}$

\vspace{0.8cm}

{\em $^{\rm a}$%
University of Iceland, Science Institute, Dunhaga 3, 107 Reykjavik, Iceland\\}

\vspace{0.3cm}

{\em $^{\rm b}$%
Laboratoire de Physique Th\'eorique, \'Ecole Normale Sup\'erieure and Institut de Physique Th\'eorique
Philippe Meyer, 24 rue Lhomond, 75231 Paris Cedex 05,
France\\}

\vspace{0.3cm}

{\em $^{\rm c}$%
Helsinki Institute of Physics, P.O.Box 64, FI-00014 University of Helsinki, Finland\\}

\vspace*{0.3cm}

{\em $^{\rm d}$%
\href{http://hep.physics.uoc.gr}{Crete Center for Theoretical Physics},
Department of Physics, University of Crete, 71003 Heraklion, Greece.
~\\
$^e$Univ Paris Diderot, Sorbonne Paris Cit\'e, \href{http://www.apc.univ-paris7.fr/APC_CS/}{APC},
UMR 7164 CNRS, F-75205 Paris, France.}
~\\

\vspace*{0.8cm}

\end{centering}

\noindent
We consider the finite temperature phase diagram of holographic QCD in the
Veneziano limit ($N_c\to\infty$, $N_f\to\infty$ with $x_f=N_f/N_c$ fixed)
and calculate one string-loop corrections to the free energy in
certain approximations. Such corrections, especially due to the pion modes 
are unsuppressed in the Veneziano limit.
We find that under some extra  assumptions the first order transition 
following from classical gravity solutions can become second order. 
If stringy asymptotics are of a special form and there are residual 
interactions it may even become of third order.
Operationally these computations imply modelling the low temperature chiral symmetry breaking phase
with a hadron gas containing $N_f^2$ massless Goldstone bosons and an exponential spectrum of
massive hadrons. A third order transition is possible only if repulsive hadron interactions via the
excluded volume effect are included.

\vfill
\noindent


\vspace*{1cm}

\noindent

\vfill

\end{titlepage}

\tableofcontents

\section{Introduction}
In this paper we consider
thermodynamics of QCD at finite temperature in a non-critical holographic model \cite{jk}
and show how one can effectively compute quantum 1-loop and even stringy corrections
by phenomenological considerations relating to the hadronic phase. We will then have
thermodynamics at all temperatures together with an analysis of possible phase
transitions.

Holographic models of QCD (V-QCD) in the Veneziano limit, 
\be
N_c\to \infty \sp N_f\to \infty\sp x_f\equiv {N_f\over N_c}~~{\rm fixed},
\ee
have been proposed and studied in \cite{jk,ajkkt,ajkkrt,Iatrakis:2014txa}, 
based on earlier proposals for pure Yang-Mills \cite{gursoy} and the identification of the chiral condensate 
as the tachyon of string theory \cite{ckp}.
In such theories for a small enough number of flavours, $x_f<x_c\simeq 4$ the low energy theory is QCD-like: 
it has chiral symmetry breaking, $N_f^2-1$ massless pions (when quarks are massless) and confinement.

The finite temperature study of V-QCD at zero baryon density and for massless quarks, \cite{ajkkt} revealed the 
possibility of one or two possible phase transitions, depending on the asymptotics of scalar potentials:
\begin{itemize}
\setlength{\itemsep}{-0.1cm}
\item[{(a)}] One phase transition. In this case the theory at a specific critical temperature 
$T_c$ jumps to a black hole solution with restored chiral symmetry.
The transition is a first order one and is at the same time a deconfinement 
and a chiral restoration transition.

\item[{(b)}] Two phase transitions. In this case the theory at a specific critical temperature 
$T_c$ jumps to a black hole solution which breaks chiral symmetry. The transition is 
first order and the theory is in a deconfined plasma phase with broken chiral symmetry. 
At a higher temperature there is a second transition, second order this time, and the 
system is described by a black hole solution with unbroken chiral symmetry.
    
\end{itemize}

The chiral phase transition above can be characterised by a chiral
condensate which is an exact order parameter for massless quarks. 
Although in the presence of quarks there is no order parameter for 
confinement, at large N$_c$, one can use the scaling of the free energy $F$ with $N_c$. 
A phase in which $F\sim {\cal O}(N_c^2)$  is a deconfined phase while a phase where 
$F\sim {\cal O}(1)$ is a confinement phase. 
In a holographic theory therefore,  any phase transition in which the system jumps 
from the $T=0$ saddle point (or the ``thermal gas solution" as it is known) to a 
black hole saddle point, is a deconfinement transition.  
\footnote{Of course this structure of two transitions
is not specific to holography, but a similar structure emerges in conventional
effective model approaches to QCD matter thermodynamics, see e.g.
\cite{Mocsy:2003qw,Kahara:2008yg,Fukushima:2008wg}.}

In \cite{ajkkrt} a finite temperature and finite density study was made for 
a V-QCD model with two phase transitions (in case (b) above). Since then further analysis has 
indicated that case (a) above is preferred when generic properties of the mesonic radial 
trajectories are imposed \cite{Arean:2012mq,arean}.
This is the version of the theory we will use in this work.

As it was already stressed in \cite{jk}, holographic theories in the Veneziano limit have more 
uncontrolled phenomena compared to theories in the standard 't Hooft limit. 
The reason is that in the open string theory sector where the fundamental degrees of 
freedom arise (quarks) as endpoints of open strings, the effective coupling constant 
is $g_s N_f$ where $g_s$ is the closed string coupling $g_s\sim {1\over N_c}$. 

In the 't Hooft limit where $N_f$ is kept fixed and of order one, $g_sN_f\sim {1\over N_c}$ 
and contributions from open string loops are suppressed. This is equivalent to the fact 
that quark loops are suppressed in the QFT.
In the Veneziano limit however, $g_s N_f\sim x_f\sim {\cal O}(1)$ and open string loops 
are unsuppressed. This can be easily checked in a simple one loop diagram that corrects 
a propagator and pions (or in general non-singlet mesons) go around the loop: the 
diagram for a single meson is suppressed by ${1\over N_c^2}$ but there are $N_f^2-1$ 
non-singlet mesons going around the loop, bringing back this contribution to become of ${\cal O}(1)$.

In this paper we will start an investigation of  such corrections. 
The brute-force method is to compute the one-loop corrections to the free-energy 
in the dual string theory, something that is in principle possible in string theory
\cite{book}. In our case however we are limited by the fact that we do not know the 
full string theory, and even if known it requires a compete solution at tree level 
in order for the full one-loop contribution to be computed. For comparison this is not yet 
known even in the best known case of holography: that of $\mathcal{N}=4$
super Yang-Mills.

We will have therefore to be more modest, and we will use physical arguments to 
isolate and compute the most important contributions in the domain $0<x_f<x_c$
to the free energy of the confined saddle point: 
the thermal gas solution. The reason is that at tree level the free-energy of the 
thermal gas solution is ${\cal O}(1)$ and therefore is neglected compared to the 
deconfined free energy that is ${\cal O}(N_c^2)$.
At one-loop however the correction, being proportional to the meson multiplicities 
is of ${\cal O}(N_f^2)$ and therefore of the same order as the deconfined free energy in the Veneziano limit.

The string one-loop calculation to the free energy in the confined phase can be 
divided into an infinite number of one-loop calculations where each of the string 
states goes around the loop. It depends only on the quadratic part of the tree 
level action and not on interactions.
If one can diagonalize the tree-level action then the result can be given in terms 
of one-loop finite temperature diagrams of particles with given tree level masses. 
For the important meson trajectories and the present holographic theory, 
this was done in \cite{Arean:2012mq,arean}.

For practical purposes we will split the non-singlet spectrum of the string theory 
in question (this is the one that gives the dominant contribution, the singlet spectrum 
contribution is down by a factor of $N_f^{-2}$) as follows:

\begin{enumerate}

\item{} The massless sector. This includes the $N_f^2-1$ massless pions.

\item{} The massive meson sector of the four main meson trajectories. This includes 
the fields generated out of the vacuum by the three important low dimension operators 
in the meson sector: the quark mass term (massive pseudoscalars and massive scalars), 
the vector current (massive vector mesons, starting with the $\rho$) and the axial 
current (massive axial vector mesons).
The masses and widths of these mesons were computed in \cite{Arean:2012mq,arean}, 
where it was shown that after the lightest 2-3 mesons the rest of the masses are 
described by linear radial trajectories of the form $m_n^2\simeq a~n$ 
with $a$ a universal computable constant.

\item{} The full string spectrum that includes an infinite number of extra fields. 
Such fields corresponds to higher spin mesons that appear at higher levels in the string 
spectrum and therefore the lightest mass of their trajectory is higher than the four 
basic trajectories. As the detailed string theory spectrum for V-QCD is not known, 
we will parametrize such a spectrum by a density of states in order to estimate 
their impact on the thermodynamics.
    
\end{enumerate}

Concretely, the above is implemented as follows in this holographic setup.
To begin with, the free energy $-f=p_q(T,\mu=0;N_c,N_f,m_q=0)$ in the QCD plasma phase is
computed so that it is normalised to
the Stefan-Boltzmann limit at $T\to\infty$. It is valid for $T_\rmi{min}<T<\infty$, were
$T_\rmi{min}$ is the temperature where the gravity solution corresponding to the QCD plasma phase becomes
thermodynamically unstable. 

The computation of the pressure $p_h(T)$ down to $T=0$ involves the stages 1-3 outlined
above. First, at stage 1 Goldstone bosons with
\be 
p_{h,{\rm{id}}}(T)/T^4=\fra{\pi^2}{90}(N_f^2-1)
\ee
are included. They arise as zero modes of the 1-loop computation. Comparing
with $p_q$ shows that a 1st order deconfining transition takes place.
Second, at stage 2 some states from the lower trajectories are included to give us $p_h(T)$.
A numerical analysis shows that they only have a small effect and the transition
still remains 1st order. Finally, at stage 3 the entire mass spectrum 
including lower Regge trajectories and with
mesonic interactions is needed.

We shall model this by
an exponential Hagedorn mass spectrum \cite{hagedorn}-\cite{gorenstein2}, see Eq.~\nr{mspect} below.
Effectively, one has a mesonic string model for hadrons; these are less
well developed for baryons but we here address the case of zero net baryon number, $\mu=0$.
The mass spectrum contains a number of parameters, which 
are strongly constrained by how $p_h(T)$ and $p_q(T)$ connect at $T_c$. 
We shall see that in this model it is easy to enforce constancy of both $\hat p\equiv p/T^4$
and of its logarithmic $T$-derivative,
the interaction measure $(\e-3p)/T^4$. The chiral transition would then be of 2nd order.
However, the second logarithmic $T$-derivative thereof would
then be discontinuous and the interaction measure would exhibit a sharp peak at $T_c$.

Our goal is to proceed
one step further and demand also the continuity of the second logarithmic derivative. This is actually quite
complicated to achieve, as the second logarithmic derivatives are naturally of opposite signs. We
find that to change the sign of the second derivative
of the hadronic pressure and to accomplish required equality one has to include interactions
in the hadron gas phase.  It is enough to include repulsive Van der Waals-type interactions caused by the
finite size of hadrons. The transition then is of third order
(as in 2-dimensional lattice SU($N\to\infty$) gauge theory \cite{grosswitten,wadia}) and actually
the shape of the interaction measure resembles that of a Tracy-Widom distribution \cite{tw,majumdar}.
With increasing $x_f$ one sees concretely how the thermodynamics approaches that in the conformal
region $x_f>x_c$.

Our analysis does not prove that the transition is of third order, just analyses a possible
framework. Universality arguments based on the epsilon expansion \cite{pisarskiwilczek} indicate a 1st order
transition for $N_f\ge3$ at fixed $N_c$. In \cite{ajkkrt} we discuss why this conclusion might not
be valid for the case studied here.
If the transition is continuous, $p(T)$ is analytic at all $T$. How the
hadronic and plasma phases then should be connected
for massive quarks is discussed in \cite{kapusta2014}. This
is excluded in the holographic approach in which the plasma pressure $p_q(T)$ is valid
only for $T>T_\rmi{min}$. Notice also that the analysis of QCD in the Veneziano limit by using weak 
coupling expansion on $S^1 \times S^3$ suggests that the transition is a crossover or of very 
high order~\cite{Hollowood:2012nr} (see also~\cite{Hollowood:2011ep}).

One obvious extension of the above is explicit chiral symmetry breaking induced by quark masses.
Then the massless Goldstone bosons would entirely disappear.

Section \ref{dual} below outlines the numerical computation of $p_q(T)$ from holography \cite{alho},
and summarizes hadron gas formulas.
Section \ref{sect:connect} discusses a second order
connection and the difficulties of a third order connection. The formalism of including
hadronic interactions via the excluded volume correction is presented in Section \ref{sect:3rd}
and results computed from this are presented in Section \ref{sect:fin} where also the $x_f$-dependence of
the results is discussed. A simple modelling of what happens if quarks are massive is contained in
Section \ref{sect:mq}.

\section{The pressure in high and low temperature phases}
\label{dual}

\subsection{High temperature pressure and vacuum spectrum from a holographic model}

To compute the pressure at temperatures above the phase transition, we apply  a model for a 
5-dimensional gravity dual of QCD matter with large $N_c$ and $N_f=x_fN_c$ originally
proposed in \cite{jk} and studied in \cite{ajkkt} and \cite{ajkkrt} (thermodynamics) and
\cite{arean} (mass spectrum). The details of the model have been thoroughly exposed in 
\cite{ajkkt, ajkkrt},
and here we only briefly recall some of the general details.

The model builds on asymptotically AdS$_5$ metric $g_{\mu\nu}$, specified by
$ds^2=b^2(z)(-f(z)dt^2+d{\bf x}^2+dz^2/ f(z))$.
In addition to gravity, the model contains bulk scalars, a dilaton $\lambda(z)$,
a tachyon $\tau(z)$,
and the scalar potential (the zeroth component of the gauge field) $\Phi(z)$.
The scalar $1/\lambda$ sources the field theory operator ${\rm Tr}F_{\mu\nu}^2$. 
The vacuum solution $\lambda(z)$ is therefore identified with the gauge theory coupling 
$N_cg^2(\mu)/(8\pi^2)$.

Furthermore, the dependence of the fields on the coordinate $z$ in this model is constrained 
to reproduce the renormalisation group flow of the dual gauge theory in the UV (i.e., for $z\to 0$). 
For the field $\lambda$ this means that
$1/\lambda(z)\approx b_0\log(\Lambda z)$ as $z\to0$, where $b_0={\textstyle{\frac{1}{3}\,}}(11-2N_f)$ 
is the one-loop coefficient of the beta function\footnote{The constraint is imposed so that the scheme 
independent two-loop running is reproduced; see \cite{ajkkt, ajkkrt} for details.}. Here we also 
see that $\Lambda$ is analogous to the scale $\Lambda_{\rm{QCD}}$. For the tachyon 
$\tau$, the UV behaviour is constrained by
\be
{\tau(z)\over{\cal L}}\approx m_qz(-\log\Lambda z)^{-{\gamma_0\over b_0}}+
\sigma z^3(-\log\Lambda z)^{{\gamma_0\over b_0}}\,,
\ee
where $\gamma_0={\textstyle{\frac{3}{2}}}$, the one-loop coefficient of the anomalous dimension 
of the mass operator in the dual field theory. Also, ${\cal L}$ is the UV AdS radius, i.e., 
$b(z)\approx {\cal L}/z$ as $z\to 0$, and $\sigma$ is proportional to the chiral condensate 
$\langle \bar q q\rangle $ with a known proportionality constant~\cite{jarvinenmassive}. 
Finally, the boundary value of the scalar potential equals the chemical potential, 
$\Phi(0)=\mu$, which we set to zero here.

In the far IR, the model is required to lead to confinement.
The modelling of the UV and IR behaviors listed above is parametrised in terms of potentials 
$V_g(\lambda)$, $V_{f0}(\lambda)$, $\kappa(\lambda)$ and $w(\lambda)$ which appear in the 
action of the five-dimensional gravity coupled with the scalars discussed above. 
To determine the high temperature pressure and the vacuum spectrum, we apply the 
results of \cite{ajkkt, ajkkrt} with the following set of potentials\footnote{Notice that 
the chosen normalization of the potentials also fixes the UV AdS radius through ${\cal L}^2=12/(12-x W_0)$.}:
\begin{align}
V_g(\lambda)&=12\,\biggl[1+{88\lambda\over27}+{4619\lambda^2
\over 729(1+2\l)}+3 e^{-1/(2\l)}(2\l)^{4/3}\sqrt{1+\log(1+2\lambda)}\biggr]\,,
\la{vg}\\
V_{f0}(\lambda)&=W_0+{8\over27}\biggl[24+(11-2x_f)W_0\biggr]\lambda
\biggr.\nn
&\phantom{=}\biggl.
+\,{1
\over 729}\biggl[24(857-46x_f)+(4619-1714x_f+92x_f^2)W_0\biggr]\lambda^2+{120\l^3\over(1+2\l)^{2/3}}\,,
\label{vf}\\
\f(\l)&={1\over \left(1+{115-16x_f\over18}\l+20\l^2\right)^{2/3}}\sqrt{1+{1\over200}\log(1+\l^2)}\,,
\la{kappa}\\
w(\l)&={\sqrt{2/3}\,\mathcal{L}^2\over \left(1+{115-16x_f\over18}\l+20\l^2\right)^{2/3}}
\left[1+{1\over200}\log(1+\l^2)\right]\,.
\la{wee}
\end{align}
The ${\cal O}(1)$, ${\cal O}(\lambda)$ and ${\cal O}(\lambda^2)$ terms are tuned so that the solutions lead to
QCD beta function and mass anomalous dimension satisfying the standard UV expansions.
The asymptotic value of $V_{f0}$ in the UV, $W_0$, remains a parameter. Its range
is $0<W_0<12/x_f$ and we have chosen $W_0=3/11$.

In the IR, at large $\lambda$, confinement (area law) and linearity of the asymptotic glueball 
trajectories at high excitation numbers require that
$V_g\sim\lambda^{4/3}\sqrt{\log\lambda}$ \cite{gursoy}. Moreover we have chosen the remaining 
parameters in~\eqref{vg} such that a good fit to YM thermodynamics is obtained, and the form 
of $V_{f0}$, $\kappa$ and $w$ at large $\l$ (as well as the value of $W_0$) such that the phase 
diagram as a function of $x_f$ is reasonable and the asymptotics of the meson spectra is linear 
with equal slopes in all sectors~\cite{arean,ajkkt}.

\begin{figure}[!t]
\begin{center}

\includegraphics[width=0.49\textwidth]{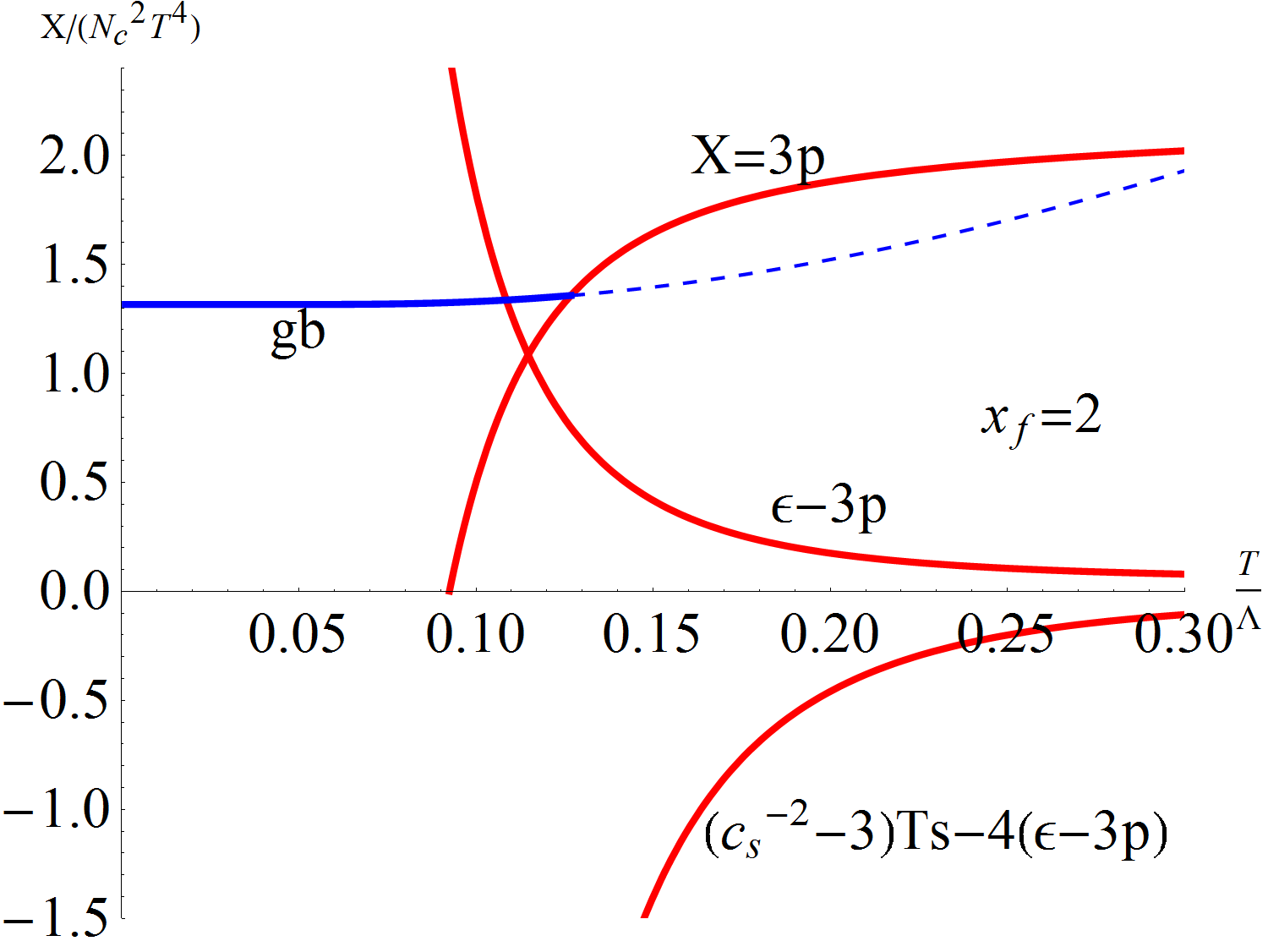}
\end{center}

\caption{\small The scaled plasma phase pressure and its first and second
$\log T$-derivatives, plotted for $x_f=2$ and for small $T$ (for the relations to
energy density $\e$ and sound velocity $c_s^2$, see \protect\ref{h1st} and \protect\ref{h2nd}).
The potentials used are those in
Eqs.\protect\nr{vg}-\protect\nr{kappa}. At $T\to\infty$ the
$3p_q$ curve approaches $3\pi^2/10=2.96$ and the zero of the pressure is at $T/\Lambda=0.0925$.
At $T\to0$ the Goldstone bosons have $3p_h/(N_c^2 T^4)=x_f^2\pi^2/30$. Without massive hadrons
there would be a 1st order transition at $T=0.125\,\Lambda$. Including the massive hadrons in
\protect\ref{vector2} and \protect\ref{scal2} gives only a marginal effect in the curve
marked gb.
}
\la{qgp}
\end{figure}

To obtain thermodynamics, one searches for black hole solutions with a horizon at $z=z_h$:
\begin{equation}
f(z_h)=0\,,\quad -f'(z_h)=4\pi T\,,\quad s={A\over4G_5}={b^3(z_h)\over 4G_5}\,.
\end{equation}
Using the code in \cite{alho} thermodynamics can be computed for a given set of model
functions. As an example, Fig.~\ref{qgp} shows for
$x_f=2$ the scaled pressure $p_q/T^4$ and its first and second derivatives with respect to $\log T$.
Note that the second derivative is negative.
The pressure is Stefan-Boltzmann-normalised, i.e., 
$p_q/(N_c^2T^4) \to (2+\fra72 x_f)\pi^2/90$ as $T \to \infty$. It is this set of curves (and
similar ones for $x_f=1,\,2.5$) that we take as the equation of state in the plasma phase.

As discussed in the introduction, this holographic setting, in addition to the thermodynamics, 
allows also the vacuum spectrum to be determined \cite{arean}. The computation involves linearising
the equations of motion around the extremal solutions of the action of the model and solving
these linearised equations of motion. In this sense it is the 1loop computation discussed in
the Introduction. For the potentials of Eqs. \protect\nr{vg}-\protect\nr{wee},
the lowest vector masses are computed to be
\begin{alignat}{6}
&2.815\,,&\ &4.731\,,&\ &6.076\,,&\ &7.188\,,&\qquad\qquad &(x_f=1)\,,&
\la{vector}\\
&0.707\,, &\ &1.156\,,&\ &1.477\,,&\ &1.744\,,&\qquad\qquad &(x_f=2)\,,&
\la{vector2}\\
&0.0795\,,&\ \ &0.1270\,,&\ \ &0.1612\,, &\ \ &0.1897\,,&\qquad\qquad &(x_f=2.5)\,,&
\la{vector2p5}
\end{alignat}
while the lowest scalar masses are
\begin{alignat}{6}
&2.640\,,&\ &4.568\,,&\ &5.857\,,&\ &6.958\,,&\qquad\qquad &(x_f=1)\,,&
\la{scal}\\
&0.599\,, &\ &1.102\,,&\ &1.405\,,&\ &1.670\,,&\qquad\qquad &(x_f=2)\,,&
\la{scal2}\\
&0.06224\,,&\ \ &0.1198\,,&\ \ &0.1519\,, &\ \ &0.1802\,,&\qquad\qquad &(x_f=2.5)\,,&
\la{scal2p5}
\end{alignat}
all in units of $\Lambda$. The axial vector and pseudoscalar masses are less relevant since
they are larger; the lowest axial vector masses are $4.289,\,1.092,\,0.1249$ and the
lowest pseudoscalar masses are $4.863,\,1.173,\,0.1279$ for $x_f=1,\,2,\,2.5$, respectively.
Including these states and their radial recurrences using formulas in the following section
gives the pressure marked gb in Fig.~\ref{qgp}. In the relevant transition region they 
have a marginal effect, the temperature in the quark phase is so low that these massive
hadronic states are hardly excited.

\subsection{The pressure at low temperature modelled as a hadron gas}
\la{sect:hg}
At $T\to0$ the relevant degrees of freedom are the Goldstone bosons spanning the coset space
SU($N_f$)$\times$SU($N_f$)$/$SU($N_f$). The pressure of these massless bosons is
\be
\frac{3p}{N_c^2T^4}=x_f^2 \frac{\pi^2}{30}\,,
\ee
which is also shown in Fig. \ref{qgp}.
This contribution alone, when matched with the high temperature contribution determined in the 
previous section, would
lead to strong first order chiral transition at $T=0.125\,\Lambda$ at $x_f=2$. 
However, also the massive hadrons are expected to contribute to thermodynamics for temperatures 
around or higher than their masses. Above we saw that the calculable low-spin masses with their
radial excitations have a negligible effect in the $T$ range relevant for phase transitions, a complete
mass spectrum, possibly also with mesonic interactions, is needed.

Generally, the ideal boson gas pressure per degree of freedom is 
\be
p_h(T,\mu,m)=\int{d^3p\over(2\pi)^3}\,{p^2\over 3E}\,{1\over e^{(E-\mu)/T}-1}={T^2m^2\over2\pi^2}
\sum_{k=1}^\infty{1\over k^2}e^{k\mu/T}K_2\left(k{m\over T}\right)\,.
\la{pid}
\ee
We will set $\mu=0$ for the rest of this paper, but we will need formulas with $\mu$ to impose
interactions via the excluded volume effect (see Eq.~\nr{pexcleq} below). 
From here one derives further
\be
T{\p\over\p T}{p_h(T,m)\over T^4}={1\over2\pi^2}\sum_{k=1}^\infty {1\over k}{m^3\over T^3}
K_1\left(k\fr{m}{T}\right)\,,
\la{ph1stder}
\ee
and
\be
\biggl( T{\p\over\p T}\biggr)^2{p_h(T,m)\over T^4}=
{1\over2\pi^2}\sum_{k=1}^\infty \left[{m^4\over T^4}K_0\left(k\fr{m}{T}\right)-
\fr2{k}{m^3\over T^3}K_1\left(k\fr{m}{T}\right)\right]\,.
\la{2ndpointline}
\ee

We want to fold the pressure with the mass spectrum
\be
\rho(m,b,a,m_0) =\delta(m)+ {\rho_0\over m_0}\biggl({m\over m_0}\biggr)^a\,e^{bm}\theta(m-m_0),
\la{mspect}
\ee
where $\rho_0$ is a dimensionless number.
The degeneracy factor $N_f^2$ of the $m=0$ Goldstone bosons and the massive flavor nonsinglet states
was factored out from~\eqref{mspect}. For the massive states with $m>m_0$, we assume an exponential 
Hagedorn spectrum together with a power of $m$. 
Using this spectrum we get for the scaled hadron gas pressure
\be
\ph(T,b,a,\rho_0,m_0)\equiv {p_\rmi{h}\over N_c^2T^4}={\pi^2\over 90}x_f^2
+{\rho_0\over m_0}x_f^2\int_{m_0}^\infty dm\,{m^a\over m_0^a} \,
e^{bm}{m^2\over 2\pi^2T^2}K_2\left({m\over T}\right)\,,
\la{phad}
\ee
where we approximated\footnote{Note that including only the $k=1$ term provides a very good approximation:
even at $m=0$ the exact result $\ph(T,0)=\pi^2/90=0.10966$ deviates just a little
from the approximate one $1/\pi^2=0.1013$.} the sum over $k$ by the first term $k=1$ 
in the contribution from massive states. Note that the
dimensionless quantity $p/T^4$ can only depend on the dimensionless combinations $T/m_0$ and $bm_0$.

We will aim at matching the 1st and 2nd logarithmic derivatives of the pressure with those of the 
high temperature phase, so we compute their expressions here. 
By using~\eqref{ph1stder} and~\eqref{2ndpointline} we find that
\be
\ph'(T,b,a)\equiv T{\partial\over\partial T}{p_\rmi{h}\over N_c^2T^4}={\epsilon-3p\over N_c^2T^4}=
{\rho_0\over m_0}x_f^2\int_{m_0}^\infty dm\,{m^a\over m_0^a} \,e^{bm}{m^3\over 2\pi^2T^3}
K_1\left({m\over T}\right)
\la{h1st}
\ee
and
\ba
\ph''(T,b,a)&\equiv& \biggl(T{\partial\over\partial T}\biggr)^2{p_\rmi{h}\over N_c^2T^4}
={(c_s^{-2}-3)(\epsilon+p)-4(\epsilon-3p)\over N_c^2T^4}\nn
&=&
{\rho_0\over m_0}x_f^2\int_{m_0}^\infty dm\,{m^a\over m_0^a} \,e^{bm}{m^3\over 2\pi^2T^3}
\biggl[{m\over T}K_0\left({m\over T}\right)-2K_1\left({m\over T}\right)\biggr]\,.
\la{h2nd}
\ea
Note how both of these vanish in the conformal case $\e=3p$ and $c_s^2=1/3$.

So far we have not specified the physical value of the unit of energy $\Lambda$.
It is the same
for thermodynamics (Fig.~\ref{qgp}) and the hadron masses (Eqs.\nr{vector}-\nr{scal2p5}).
It could be fixed in GeV units if one, e.g., knew $T_c$ in GeV units for $x_f=1$. Its $x_f$
dependence requires further study.

The integrals in \nr{phad}-\nr{h2nd}, of course, blow up for $T>T_\rmi{Hagedorn}=1/b$.
However, exactly at $T=1/b$ the integrand at large $m$ is $\sim m^{a+3/2+i}$,
where $i$ is the order of the derivative, so that the $m$-integral converges for sufficiently
negative $a$. The hadron gas pressure then does not diverge but approaches a constant
as $T \to 1/b$ from below. This will be the case in practice.

To further elaborate on the functional form of the spectral weight $\rho(m)$, we note first that
experimental hadron data cannot fix the form of the spectral weight. Simply, the available
range of masses is too small to, e.g., separate a power from exponential. For example,
over the range $1<m/\gev<2$ numerically \cite{broniowski3,cleymans}
\be
4.52\exp(2.76\,m/\gev)\approx {0.48\over((m/\gev)^2+0.25)^{5/2}}\exp(5.75\,m/\gev)\,.
\ee

We also note that a large number of papers have been written on $\rho(m)$, starting from the classics by
Hagedorn \cite{hagedorn}, Huang-Weinberg \cite{huangweinberg} and Frautschi \cite{frautschi}.
Some examples are as follows:
Ref. \cite{kapusta} computes $\rho(m)$ in the bag model,
\cite{dienes} writes ``confining phase is consistent with effective string theory in which
conformal symmetry and modular invariance play a significant role", \cite{freund} for the first
time discusses a separate density for mesons and baryons, but believes that they should
be equal to exponential accuracy, \cite{leonidov} tries to connect hadronic and plasma
phases like here, \cite{broniowski1}-\cite{broniowski3} believes that the mesonic and baryonic densities
should be different and fits both of them, \cite{cohen} doubts the empirical validity of
the exponential mass spectrum and \cite{cleymans} updates mass spectrum fits and
includes the chemical potential in the hadron gas discussion.
From all the work on $\rho(m)$ it is obvious that there is no unique parametrisation for
it.

\section{Connecting hadron gas with plasma}
\la{sect:connect}
We now keep fixed the plasma phase thermodynamics, plotted in Fig.~\ref{qgp} for $x_f=2$.
The solution is chirally symmetric, i.e., corresponds to the zero value of the bulk tachyon 
$\tau$. The pressure of the plasma phase vanishes at
$T/\Lambda=0.0925$ for $x_f=2$. In the hadron phase the pressure is given in \nr{phad} and
depends on a number of parameters. The question then is how QCD dynamics connects
these two curves and what this implies for the properties of the hadron phase.

We shall attempt to make the transition as continuous as possible. For this one
needs to satisfy the matching conditions, in increasing order:
\bi
\item 1st order transition: only $\hat p$ continuous,
\item 2nd order transition: Both $\hat p$ and $\hat p'$ continuous,

\item 3rd order transition:  $\hat p$, $\hat p'$ and $\hat p''$ continuous.
\ei

To begin with, in the hadronic phase at low $T$ one at least has
massless Goldstone bosons, which concretely arise due to chiral symmetry breaking.
Their $T\to0$
contribution to $3 \hat p_h$, $x_f^2\pi^2/30$, is also plotted in Fig.~\ref{qgp}, together with
a marginal effect of the lowest calculable masses.
The curve denoted by ``gb'' in the figure simply continues to the plasma
line and gives a 1st order chiral (and deconfining) transition when one moves from
one pressure curve to the other. The transition temperature would be
at $T_\rmi{gb}=0.125\,\Lambda$, somewhat higher than $0.0925\,\Lambda$ 
(where the pressure of the plasma phase goes to zero).

The model formally has 5 parameters: $m_0,b,a,\rho_0$ and\footnote{With the understanding that also the 
continuity of pressure is a constraint for the parameters. Then also the critical temperature, i.e., 
the temperature where the constraints are evaluated, is a free parameter.} the value of $T_c$. The
minimum mass $m_0$ can be determined holographically by a separate computation,
see Eqs. \nr{vector}-\nr{scal2p5}. This
is conceivably possible also for the Hagedorn temperature $1/b$, but this has not yet been done.
Further, $a$ is expected
to be negative and $T_c$ has to be somewhat above the point at which $p_q$ vanishes
in Fig.~\ref{qgp}. Even though the number of parameters is in principle high enough to 
match all conditions, it is not
clear that a solution exists. In fact, we shall see that a third order solution is found only if
interactions in the hadron gas are taken into account. This happens \`a la Van der Waals by including the
effects of finite size of hadrons.

Even if we ultimately determine the parameters $a$, $\rho_0$ and $T_c$ by numerical fitting, it is useful to
have a simple toy model to estimate their values. To obtain the functional form of
Eq. (\ref{mspect}), consider the eigenvalues $N$ of the operator
$\sum_{n=1}^\infty\sum_{\mu=1}^{d-2}nN_{\mu n}$.
The degeneracy of the eigenvalue is ($d-2$ is the number of transverse dimensions) \cite{book}
\be
P(N,d)={1\over\sqrt2}\biggl({d-2\over24}\biggr)^{(d-1)/4}N^{-{d+1\over4}}
\exp\left(2\pi\sqrt{{d-2\over6}N}\right)\,.
\ee
We change this to mass density by $P(N,d)dN=\rho(m)dm$, $N=\alpha' m^2$:
\be
\rho(m)=\sqrt2\,\sqrt{\alpha'}\biggl({d-2\over24}\biggr)^{{d-1\over4}}(\sqrt{\alpha'}m)^{-{d-1\over2}}
\exp\left(2\sqrt{{d-2\over6}}\pi\sqrt{\alpha'}m\right)\,.
\ee
Choosing $d=5$ and $m_0=1/\sqrt{\alpha'}$ this is of the form \nr{mspect} with
\be
\rho_0={\sqrt2\over8} \approx 0.177\,,\quad a=-2\,,\quad b m_0=\pi\sqrt2 \approx 4.44\,.
\la{modpar}
\ee
Our final 3rd order numbers will not agree with this toy model.

\begin{figure}[!t]
\begin{center}

\includegraphics[width=0.49\textwidth]{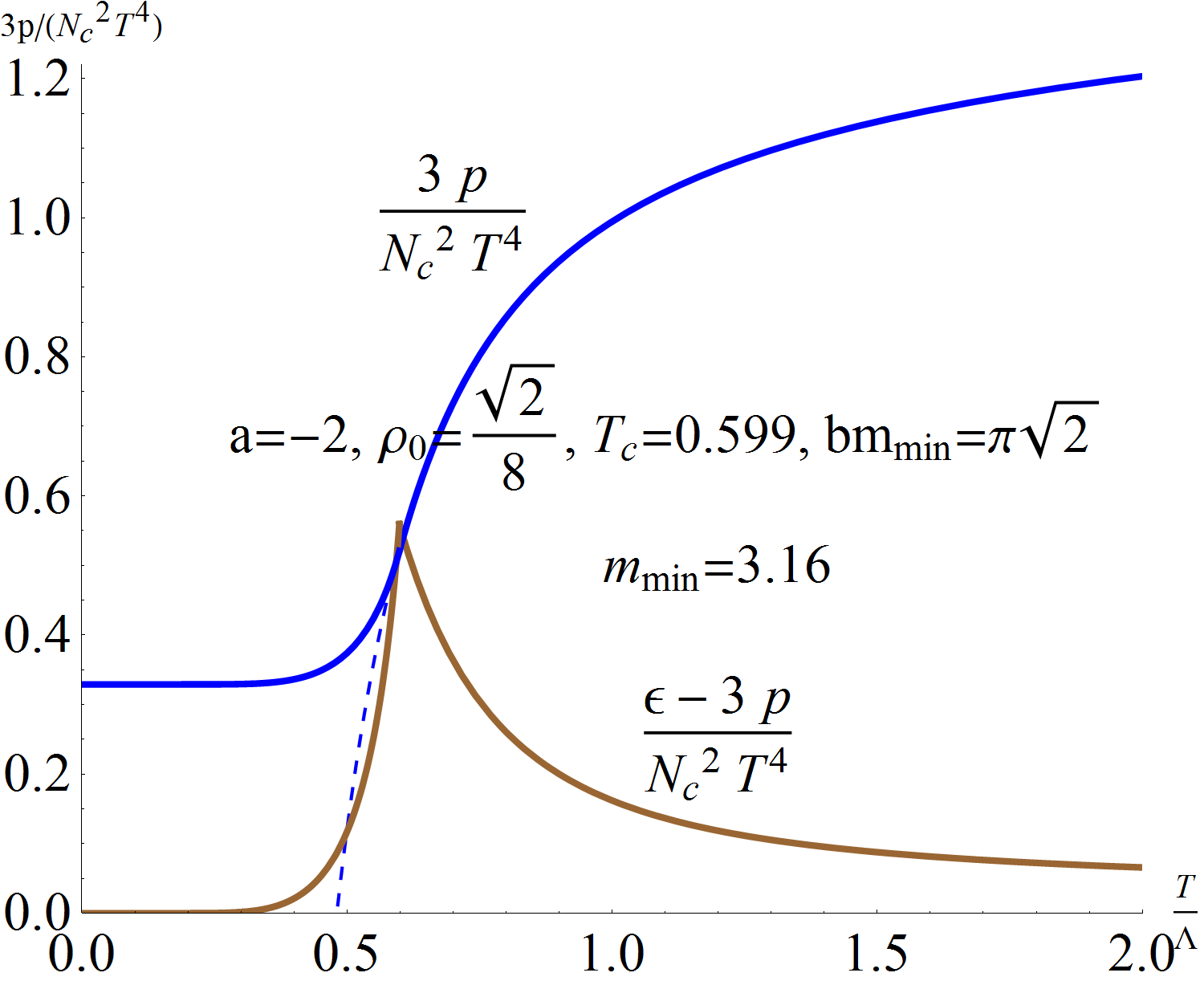}
\includegraphics[width=0.49\textwidth]{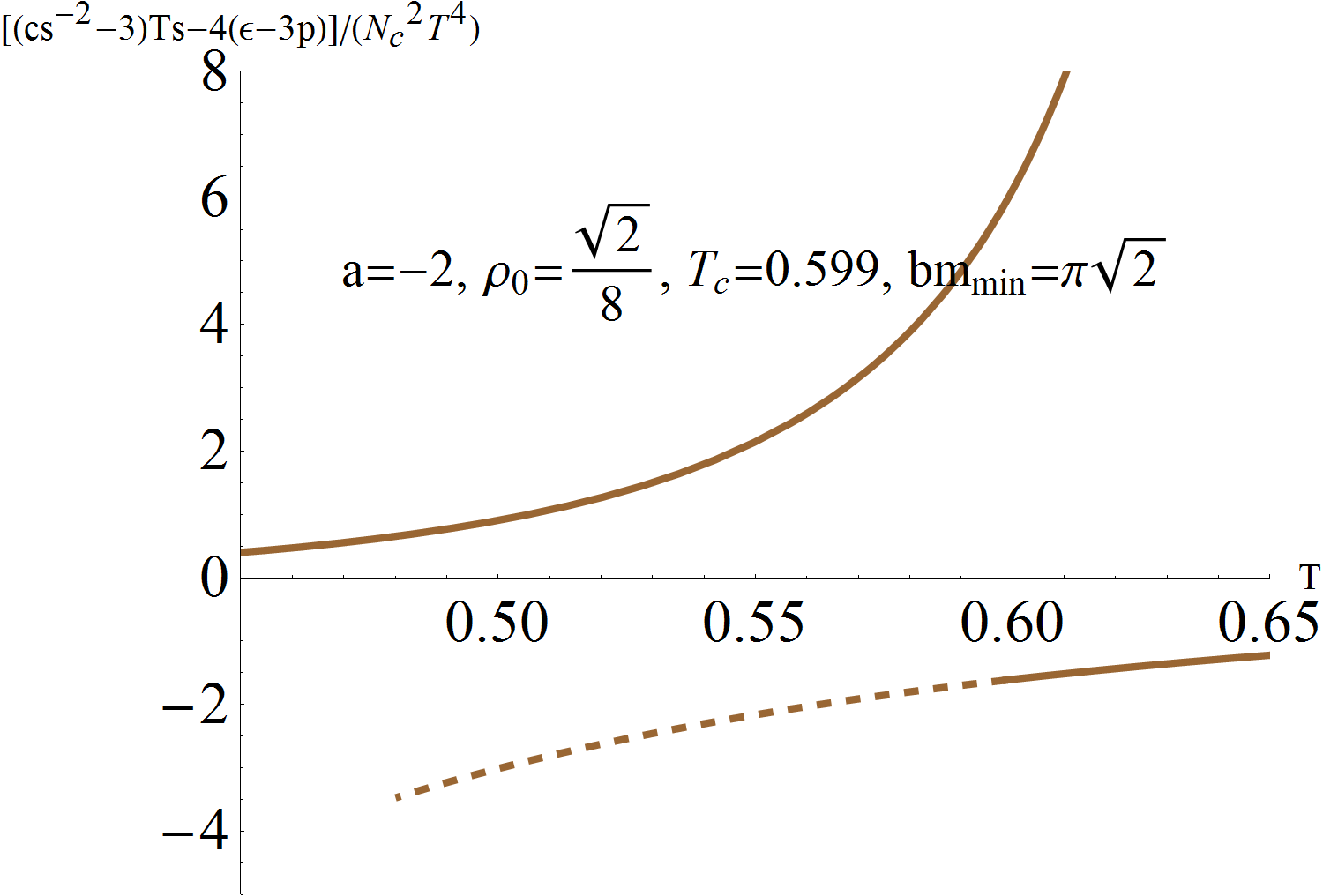}
\end{center}

\caption{\small Left: Thermodynamics at $x_f=1$ with the toy model parameters in
\protect\nr{modpar} and the fitted values $T_c=0.5987$, $m_0\equiv m_\rmi{min}=3.158$. Right: The
2nd derivatives as functions of $T$. Note the opposite signs, corresponding to the sharp peak in the
interaction measure in the left panel. }
\la{f2}
\end{figure}

\subsection{Second order transition}
A second order transition, i.e., a transition
where $\hat p$ and $\hat p'$ are continuous, is very easy to obtain.
As an example,
consider the model \nr{modpar}. Then one has two quantities to determine: the
minimum mass $m_0$ and the transition temperature $T_c$. One first  determines $m_0=m_0(T)$ numerically
by requiring continuity of $\hat p'$. Then
inserting this to the continuity condition for
 $\hat p$, one determines the value\footnote{The numerical values of dimensionful quantities here and 
below are given in units of $\Lambda$, unless stated otherwise.} of $T_c=0.5987$, giving finally 
$m_0=m_0(T_c)=3.158$.
The thermodynamics so obtained is plotted in Fig.~\ref{f2}.
One observes that the minimum mass obtained from the
thermal fit is close to the directly determined one in \nr{vector} but somewhat bigger.

To have an idea of the range of acceptable parameter values,
take the computed value $m_0=2.8$ from \nr{vector}. Assume $a$ has some fixed value.
Then we have 3 parameters to determine,
$T_c,\,b,\,\rho_0$, but only two equations, equality of $p/T^4$ and its $\log T$-derivative.
By demanding that the derivatives coincide one first determines $b=b(T_c,\rho_0)$. The
equation $p_h(T_c,b(T_c,\rho_0),\rho_0)=p_q(T_c)$ then gives $T_c=T_c(\rho_0)$
and finally $b=b(T_c(\rho_0),\rho_0)=b(\rho_0)$. The outcomes for $T_c$ and $b$ are plotted in 
Fig.~\ref{afix} as functions of the normalisation of mass spectrum $\rho_0$
and for various fixed values of $a$.

\begin{figure}[!t]
\begin{center}

\includegraphics[width=0.49\textwidth]{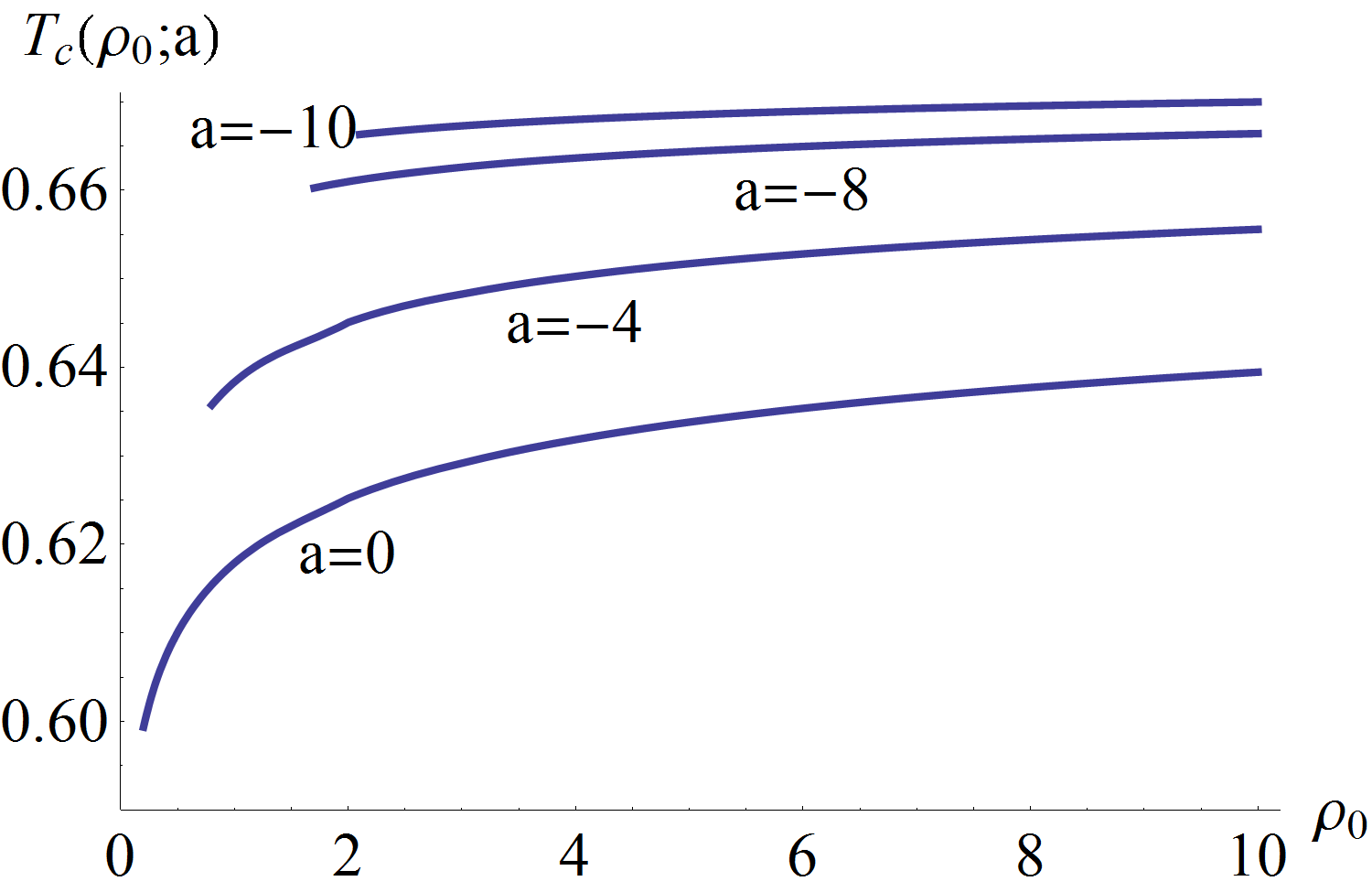}\hfill
\includegraphics[width=0.49\textwidth]{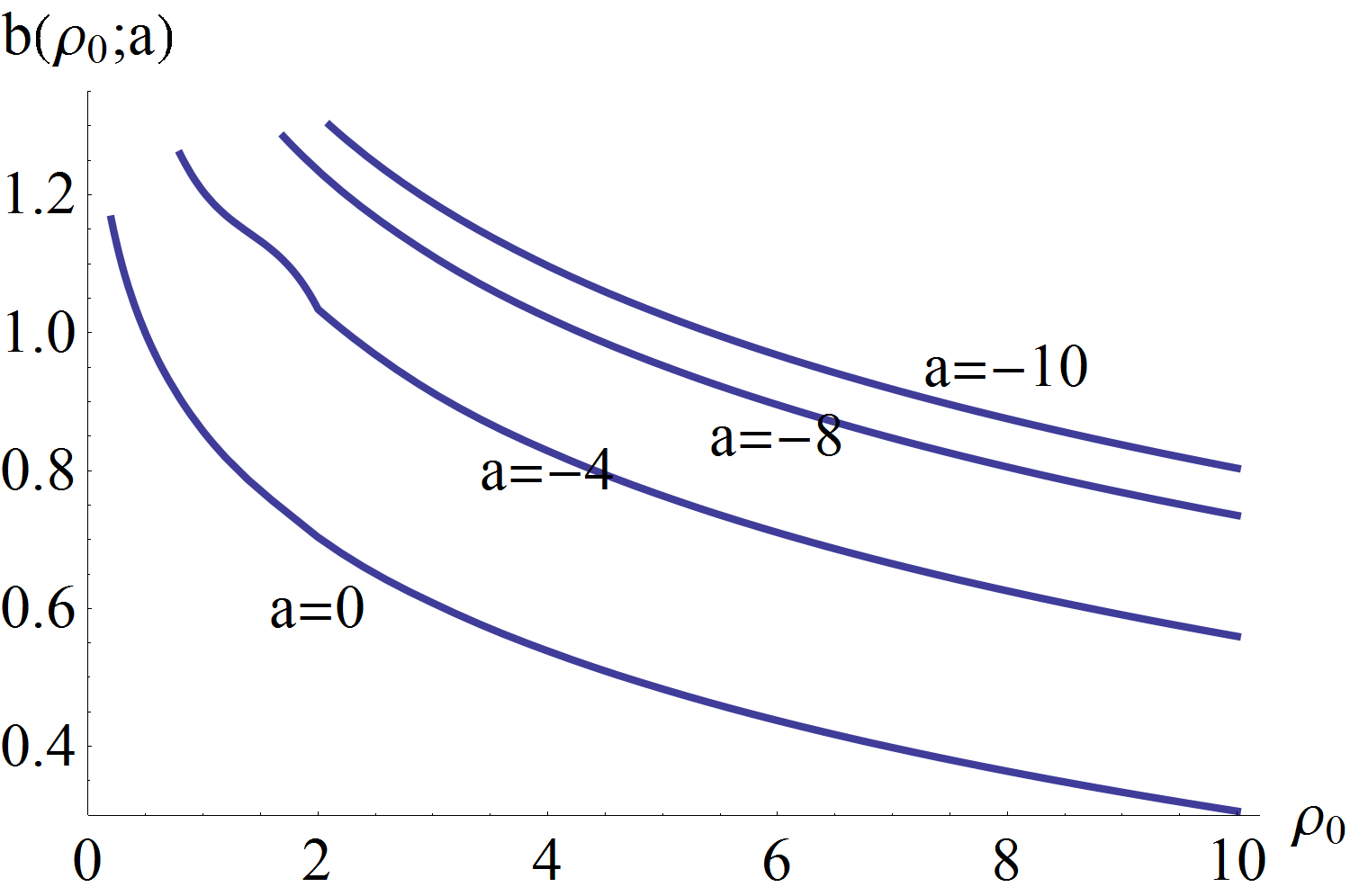}

\end{center}

\caption{\small Fitted values of $T_c(\rho_0;a)$ (left) and $b(\rho_0;a)$ (right) for $x_f=1$ and
$m_0=2.8$.}
\la{afix}
\end{figure}

\subsection{Third order transition with pointlike mesons}
The next stage is to also require that the 2nd derivatives be equal when we move
from $p_h$ to $p_q$. Since pressure in the plasma phase is taken to be fixed and $\hat p_q''< 0$,
one also has to change the sign of $\hat p_h''$ from positive to negative, 
see the right hand plot in Fig~\ref{f2}. If the second derivatives are equal and negative at some 
$T_c$, $\ph''$ must have a zero somewhere. At the same
point the interaction measure $\ph'$ will have an extremum and the sharp peak
has disappeared.

To see the sign of $\ph''(T,b,a)$ in \nr{h2nd} one can write it in the form
\be
\ph''(T,b,a)={\rho_0\over 2\pi^2}{T^{a+1}\over m_0^{a+1}}\int_{m_0/T}^\infty dy\,y^{a+3}\,e^{bT y}y^3
\left[yK_0(y)-2K_1(y)\right]\,.
\ee
The sign is determined by the combination $yK_0(y)-2K_1(y)$ which has a zero at $y=2.3864$.
At small $y$ this combination is negative and behaves approximately as $-2/y-2y\log(y)$, 
while at large $y$ it
is positive and behaves approximately as $e^{-y}\sqrt{\pi y/2}$.
Thus the lower limit $m_0/T$ allows only positive values if $T/m_0<1/2.3864=0.419$. To have a desired
negative value the integral must probe the negative region below $y=2.3864$
by having $T>0.419\,m_0$.
Not much of it is needed as seen from the numerical plots in Fig.~\ref{hder} (where $m_0=1$),
$\ph''(T,b,a)<0$ for $T\gsim 0.5$ almost independent of the values of $b$ and $a$.
At this temperature then first derivative, i.e., scaled $\epsilon-3p$ has a maximum and
the sharp peak in the interaction measure has disappeared.

For $x_f=1$ one has $m_0\approx2.8$
and one can expect $\ph''$ to become negative only for $T>0.5\,m_0\approx1.4$.
This would push the maximum of the interaction measure to values of $T$ much larger
than those encountered earlier, which are about $0.6$. It is physically obvious that
the hadron gas phase cannot be thermodynamically stable up to such
large $T$. One concludes that with the present hadron gas model, pointlike hadrons,
the sign of $\ph''$ cannot be changed, the cusp in the interaction measure
cannot be removed and the transition cannot be made of higher than 2nd order.

\begin{figure}[!t]
\begin{center}

\includegraphics[width=0.49\textwidth]{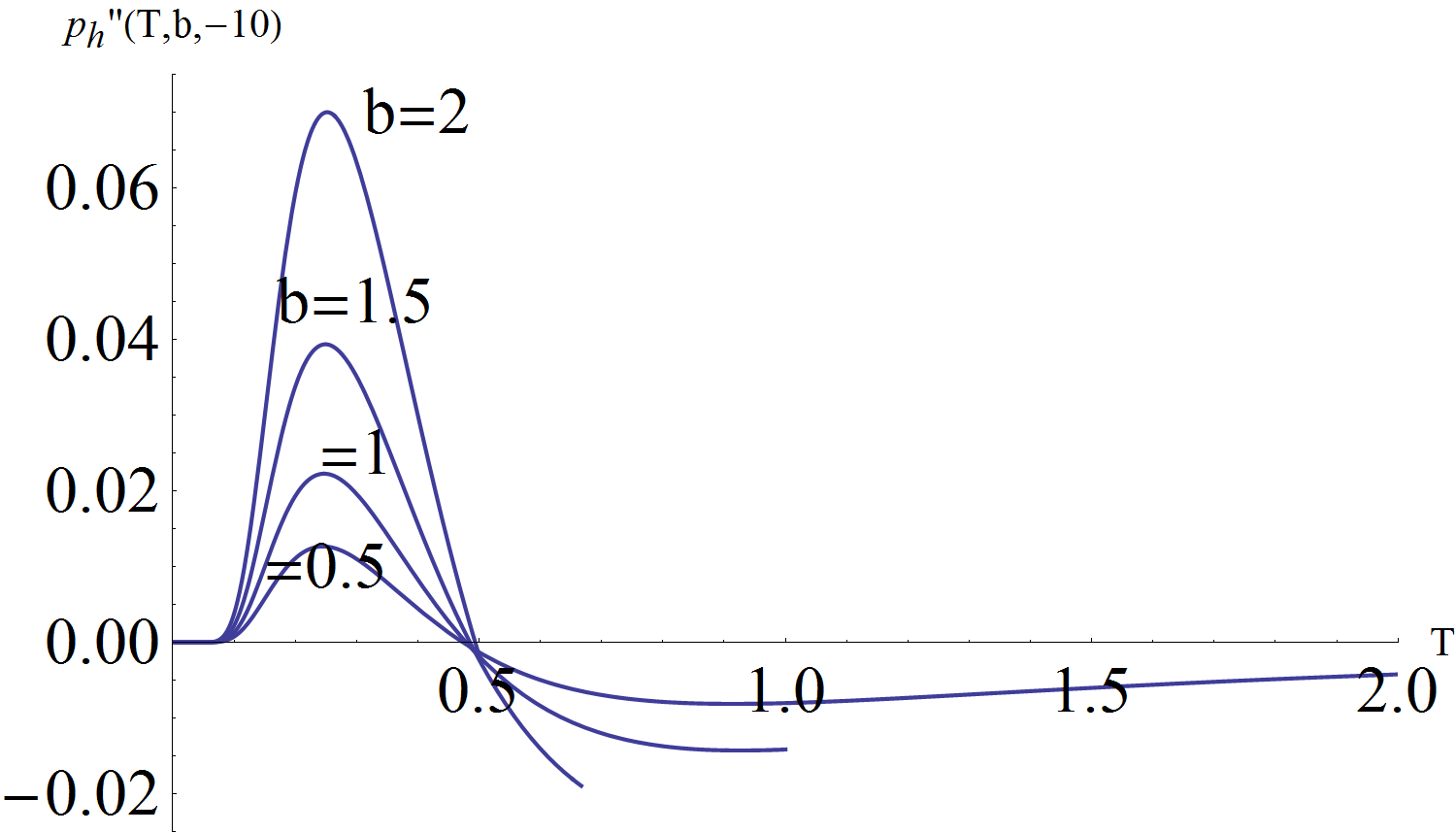}
\includegraphics[width=0.49\textwidth]{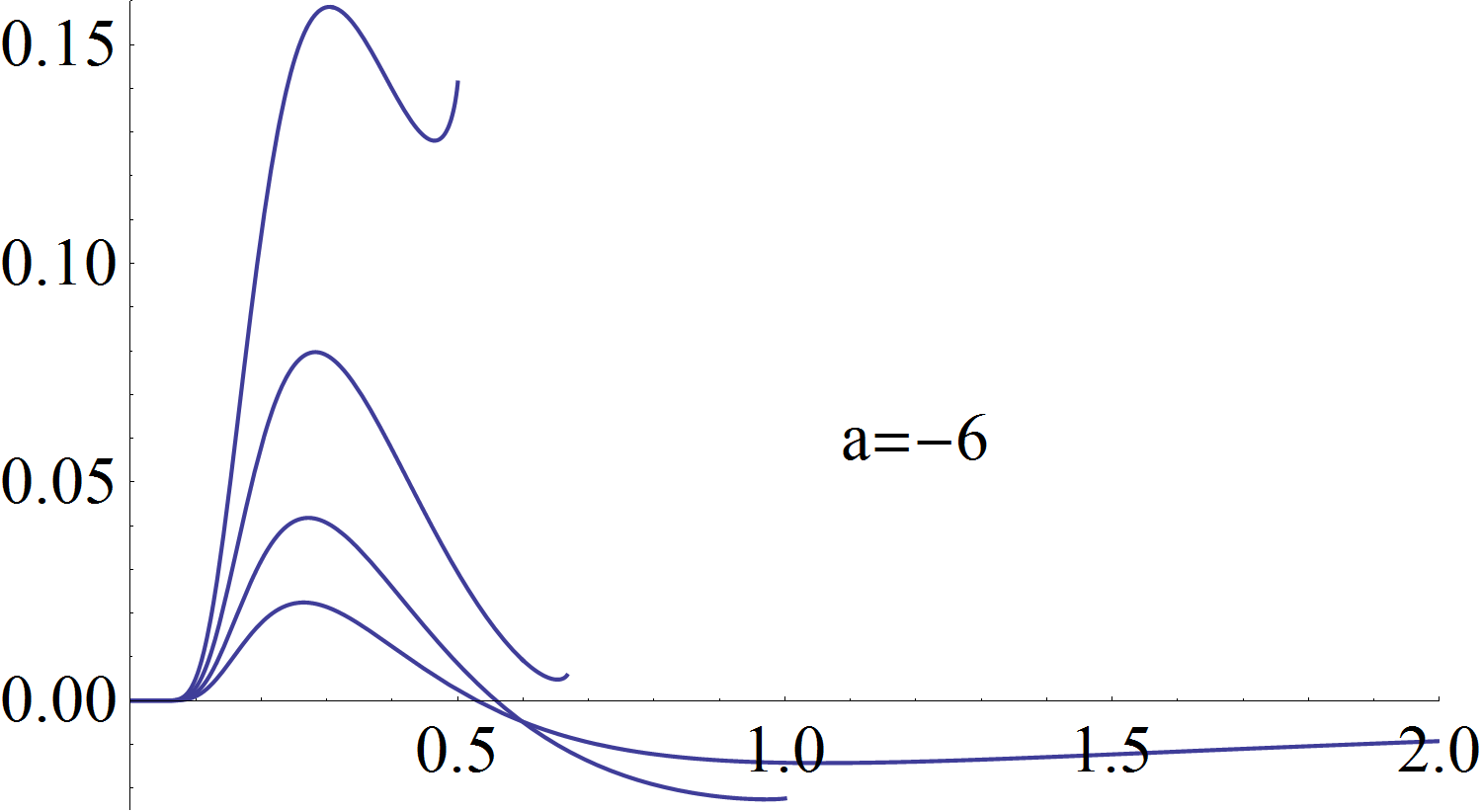}
\end{center}

\caption{\small The second derivative in \protect\nr{h2nd} with $m_0=\rho_0=x_f=1$,
evaluated for $a=-10$ and $a=-6$ and (curves from bottom) $b=0.5,\,1,\,1.5,\,2$. For $b\gsim2$ the
negative part disappears. The curves end at $T=1/b$.
The zero of $\ph''(T,b,a)$, if any, is at $T\approx 0.5$ largely independent
of the values of $b,\,a$. }
\la{hder}
\end{figure}

\section{Third order transition with excluded volume corrections}
\la{sect:3rd}
Let us then check if including hadron interactions via the excluded volume correction
\cite{kapustaolive,gorenstein,kapusta2014}
would make it possible to make the transition of third order and to
get rid of the cusp in the interaction measure. One replaces $V\to V-v_0N$ where $v_0\equiv1/T_0^3$ is
the volume of a single meson and $N$ the number of mesons.
What is the new pressure $p(T,\mu)$, given the
pointlike boson pressure $p_0(T,\mu)$ in \nr{pid}? Here $\mu$ is the chemical potential associated
with the meson number $N$; i.e. $n=N/V=\p p/\p\mu$. The chemical potential associated
with conserved baryon number is taken to be zero.

According to a related model in \cite{kapusta2014} the effective hadron volume is 
$\fra{16}{3}\pi r_h^3=(7.93/{\rm GeV})^3
=1/T_0^3$ so that $T_0=0.126\,{\rm GeV}$. In QCD $T_c\approx 0.15$~GeV while in our units
$T_c/\Lambda \approx 0.5$ (where we reinstated the unit of energy $\Lambda$). 
Thus $\Lambda \approx 2\,T_c \approx 0.3$~GeV so
that an expected magnitude at $x_f=1$ is $T_0/\Lambda\approx 0.42$. We shall find below 
by matching the pressures of the two phases for $x_f=1$ that $T_c/T_0\approx3$, $T_0/\Lambda\approx0.25$.

As shown in detail in \cite{gorenstein}, Section II, $p(T,\mu)$ is obtained\footnote{In this section,
$p_0$ is the pressure of pointlike hadron gas, $p$ is the pressure when excluded
volume effects are included.} as a solution of the
transcendental equation
\be
p(T,\mu)=p_0(T,\mu-\fra1{T_0^3}p(T,\mu))\,.
\la{pexcleq}
\ee
Taking the partial derivative with respect to $\mu$ of \nr{pexcleq} one obtains for the number density
\be
n(T,\mu)={n_0(T,\mu-\fra1{T_0^3}p(T,\mu))\over 1+\fra1{T_0^3}n_0(T,\mu-\fra1{T_0^3}p(T,\mu))}\,.
\la{nex}
\ee
Thus, to apply this, one has to solve $p(T,\mu)$ from \nr{pexcleq}.

In the Maxwell-Boltzmann (MB) approximation  $e^{\beta (E-\mu)}\gg1$, and only the $k=1$ term
in the series \nr{pid} contributes,
\be
p_0(T,\mu)={m^2T^2\over2\pi^2}K_2(\fra{m}{T})e^{\mu/T}=Tn_0(T,\mu)\,.
\la{p0MB}
\ee
The $\mu$ dependence is a simple exponential and  \nr{pexcleq} becomes
\be
{p(T,\mu)\over p_0(T,\mu)}=\exp\biggl(-{p_0(T,\mu)\over T T_0^3}{p(T,\mu)\over p_0(T,\mu)}\biggr)=
\exp\biggl(-{n_0(T,\mu)\over  T_0^3}
{p(T,\mu)\over p_0(T,\mu)}\biggr)\,.
\la{MBexcl}
\ee
In our case $\mu=0$, and inserting this in \nr{MBexcl} we see that all quantities are just
functions of $T$.
The equation is of the general form $q=e^{-a q}$ which is trivial to solve numerically. More formally,
the solution is
\be
q(a)={1\over a}W(a)
\la{qa}
\ee
where $W(a)$ is the Lambert's function, ProductLog in Mathematica parlance.
At small $a$
\be
q(a)={1\over a}W(a) = \sum_{k=0}^\infty (-1)^k{(k+1)^{k-1}\over k!} a^k=1-a+\fra32 a^2-\fra83 a^3+ \cdots \, ,
\ee
and at large $a$
\be
q(a)={1\over a}W(a)={1\over a}(\log a-\log\log a+\CO(1))\,.
\ee
Thus the pressure after excluded volume corrections is
\be
p(T)\equiv p(T,m,T_0)=T\,T_0^3W\biggl({p_0(T)\over TT_0^3}\biggr)=
TT_0^3\,W\biggl({m^2T\over2\pi^2T_0^3}K_2\left(\fr{m}{T}\right)\biggr)\,.
\la{psoln}
\ee
At small $T/m$, $n_0(T)\sim\exp(-m/T)$ is small and
\be
p(T)=p_0(T)\biggl(1-{n_0(T)\over T_0^3}+\cdots\biggr)
\ee
and at large $T$
\be
p(T)=T\,T_0^3\biggl(3\log \fr{T}{T_0}-\log\log \fr{T}{T_0}+\CO(1)\biggr)\,.
\la{pexlargeT}
\ee

It is also illuminating to evaluate the effect of volume exclusion on the number
density at large $T$. In \nr{nex} there is an additional suppression factor
$\exp(-p/(TT_0^3))=\CO(1)T_0^3/T^3$, where we used~\eqref{MBexcl} and~\eqref{pexlargeT} to obtain the estimate.
Therefore the excluded volume density at large $T$ is simply
$n=\CO(1)T_0^3$, the hadrons are densely packed but do not overlap. Related to this,
in the excluded volume model the effective chemical potential
$-p(T,0)/T_0^3$ is always negative and one does not get to the Bose-Einstein condensation
domain $\mu\to m$ where meson wave functions overlap.

\begin{figure}[!tb]
\begin{center}

\includegraphics[width=0.49\textwidth]{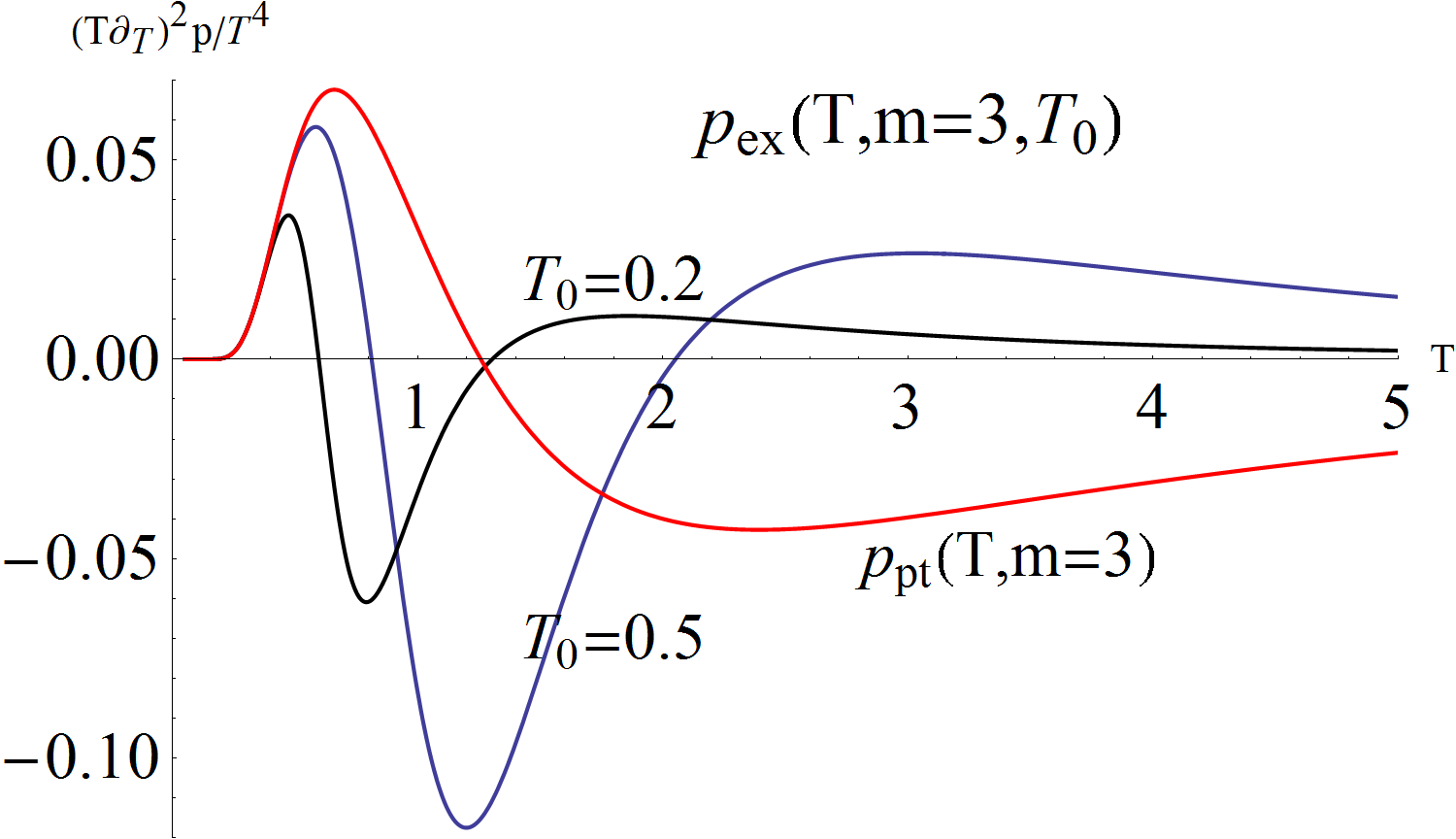}
\includegraphics[width=0.49\textwidth]{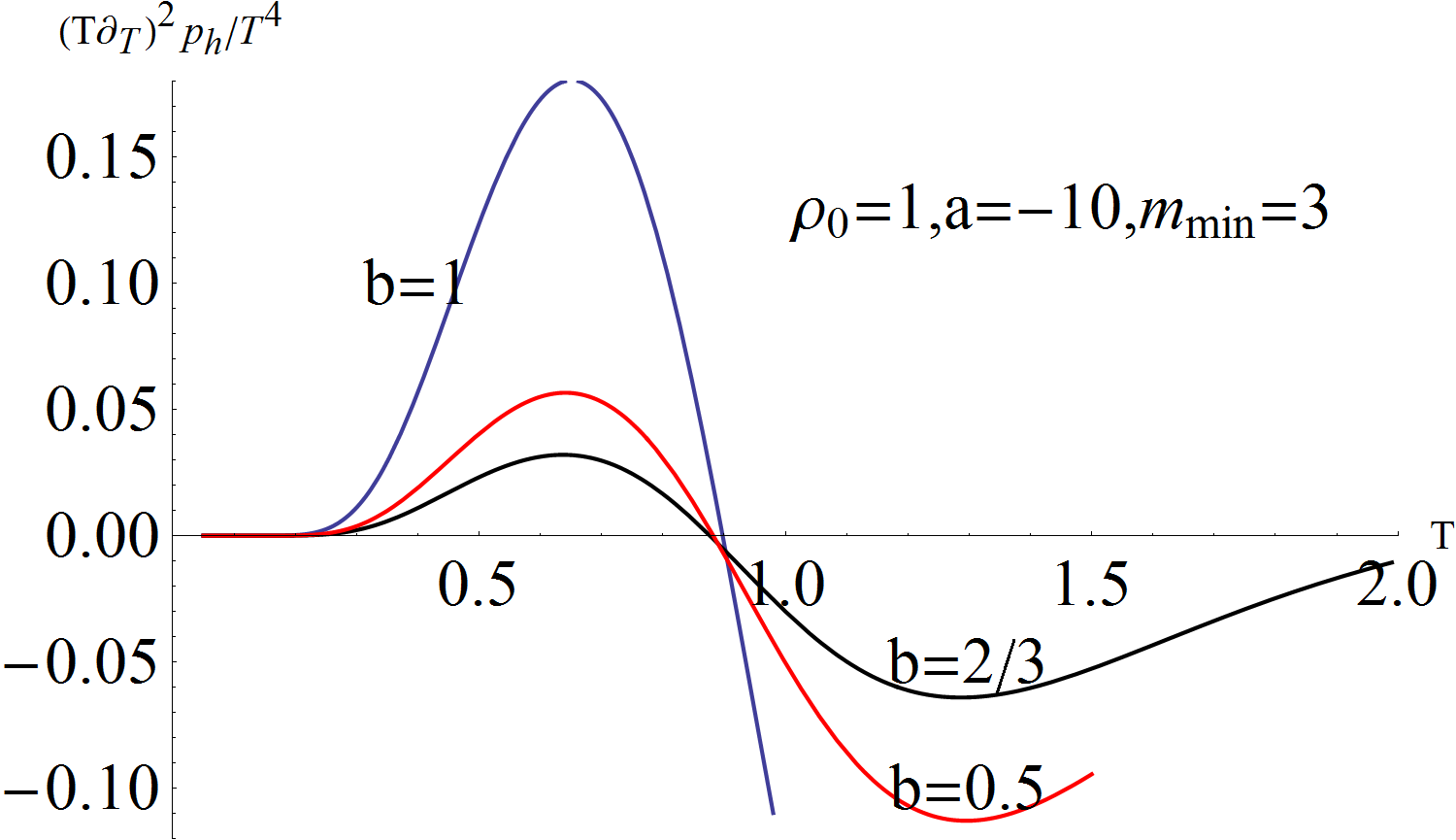}

\end{center}

\caption{\small Left: The second derivative $(T\p_T)^2p/T^4$ of the excluded volume solution
\protect\nr{psoln} for $m=3$ and two values of the meson volume parameter $T_0$. The pointlike
curve \protect\nr{2ndpointline} (with $k=1$) is also shown. Right:
The $\rho(m)$-integral of the second derivative of the excluded volume solution
\protect\nr{psoln} of \protect\nr{MBexcl} with $\rho_0=x_f=1$,
$a=-10$, $m_0 \equiv m_\rmi{min}=3$, $T_0 = 0.5$, and (curves from bottom) $b=0.5,\,2/3,\,1$.
}
\la{hderex}
\end{figure}

As a first step towards understanding the effects of excluded volume,
the left panel of Fig.~\ref{hderex} shows numerical results for
the 2nd $\log T$ derivative of the scaled pressure solution \nr{psoln} of \nr{pexcleq}
for $m=3$. There the red curve corresponds to the pointlike
pressure, the $k=1$ term in \nr{2ndpointline}. As discussed above, the 2nd derivative
is negative for $T>m/2.386=1.257$. The two excluded volume curves correspond to
meson volume parameter values $T_0=0.2$ and $0.5$. Here $0.5$ is chosen so that the hadron
volume is of the order of $1/T_c^3$ with $T_c$ as in previous figures. The small $T$ behavior is
always like in the pointlike case, the large $T$ behavior shows the positive 2nd derivative
following from \nr{pexlargeT} and in between there is a negative region. Increasing $T_0$
would extend the negative region to the right so that it asymptotically approaches
the pointlike curve. In any case meson finite size enhances the negative region.

As a second step,
the right panel of Fig.~\ref{hderex} shows numerical results for
the 2nd $\log T$ derivative of the scaled
pressure solution \nr{psoln} of \nr{pexcleq} at $T_0=0.5$ but now integrated over the
exponential mass spectrum \nr{mspect} $\rho(m,b,a,m_0)$. A large negative
value of $a$ is needed, here $a=-10$. Since we have an exponential mass
spectrum, the computation can only be valid up to $T=1/b$. Since plasma
extends down to $T=0.5$, we certainly must have $b<2$.

Comparing with Fig.~\ref{hder} one sees that indeed excluded volume based
interactions increase the magnitude of the negative 2nd derivative even when
integrated over exponential mass spectrum.
The parameter $\rho_0$ can be used to increase the magnitude
further.

Above in~\eqref{p0MB} we used the Maxwell-Boltzmann approximation which, 
due to its explicit $\mu$ dependence, led to simplified computations.
Without invoking the MB approximation, instead of $q=e^{-aq}$, one has to solve the equation
\be
q={\sum_1^\infty \fra1{k^2} e^{-kaq}K_2(k\fra{m}{T})\over \sum_1^\infty \fra1{k^2}K_2(k\fra{m}{T})}
\ee
for $q=q(a,m/T)$. One can numerically check that this improvement
has only a marginal effect. The physics reason for this is that, as discussed above, in the excluded
volume model the meson wave packets are densely packed but do not overlap, and one is never
close to Bose-Einstein condensation.

Another calculable interaction term is that among Goldstone bosons. Including terms of order 2 and 4 in
the chiral Lagrangian and to 3 loops \cite{gerberleutwyler,gomeznicola}
one has
\be
p_\rmi{gb}(T)=N_f^2{\pi^2\over90}T^4\biggl(1+{N_f^2T^4\over 144f_\pi^4}\log{\Lambda_p\over T}+\CO(T^6)\biggr)\,,
\ee
where the scale $\Lambda_p$ depends on the higher order couplings of the bosons. 
For the interaction measure we find
\be
{\epsilon_\rmi{gb}-3p_\rmi{gb}\over N_c^2T^4}={\pi^2\over30}x_f^2{N_f^2T^4\over 108f_\pi^4}
\biggl(\log{\Lambda_p\over T}-\fr14\biggr)\,.
\ee
In the Gell-Mann-Oakes-Renner relation (for a theory with all quark masses equal)
$m_\pi^2f_\pi^2=-\fra{2}{N_f}m_q\langle\bar qq\rangle$
one has $\langle\bar qq\rangle\sim N_fN_c$
so that $f_\pi$ scales as $f_\pi^2\sim N_c$. Thus in the interaction measure above $N_f^2/f_\pi^4\sim x_f^2$
and it seems that this term will be
dominated by effects from the massive states. As a side remark, it is also repulsive for
$T\lsim\Lambda_p$, it increases $p_\rmi{gb}$ there.

\begin{figure}[!tb]
\begin{center}

\includegraphics[width=0.49\textwidth]{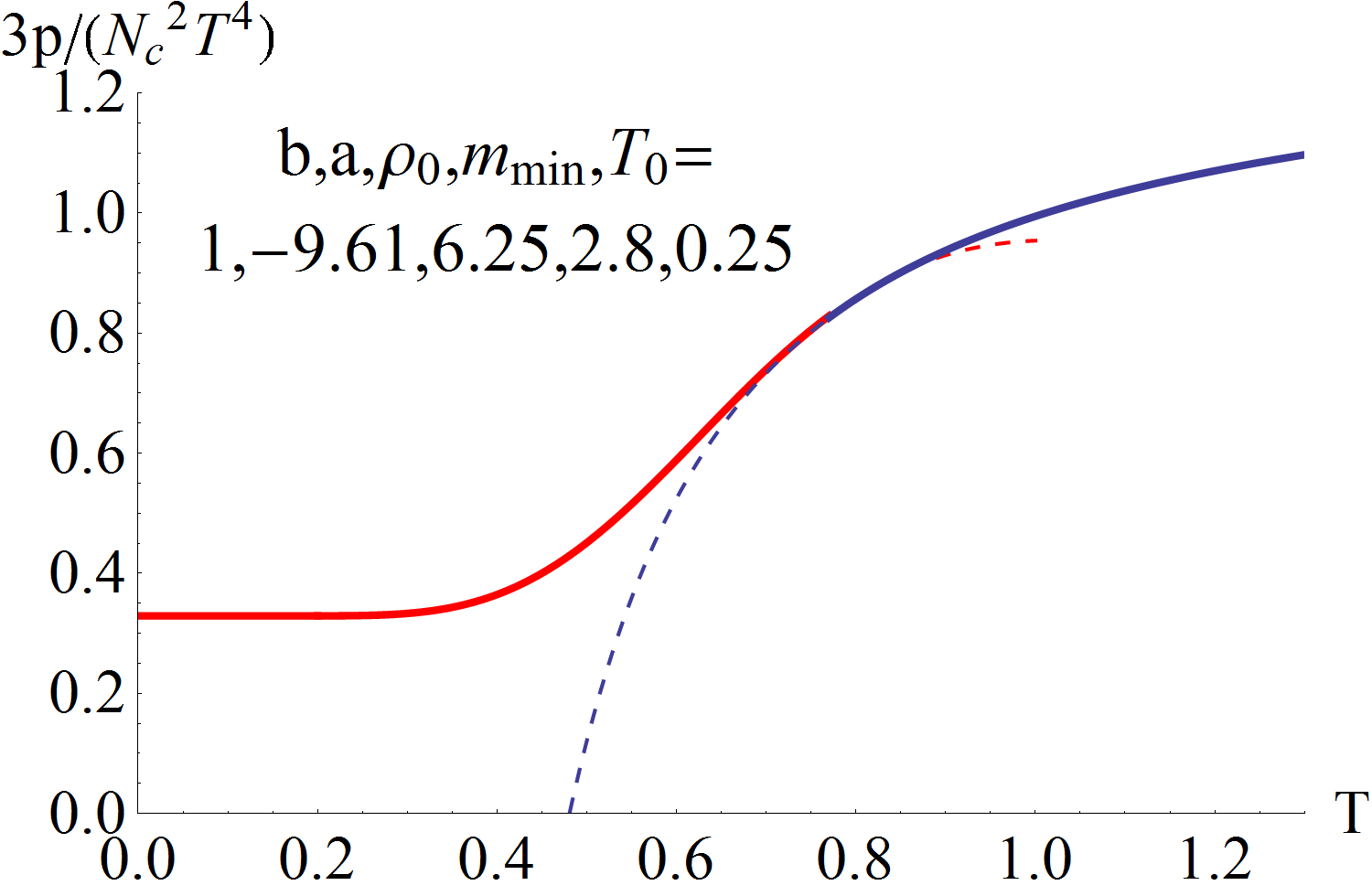}\hfill
\includegraphics[width=0.49\textwidth]{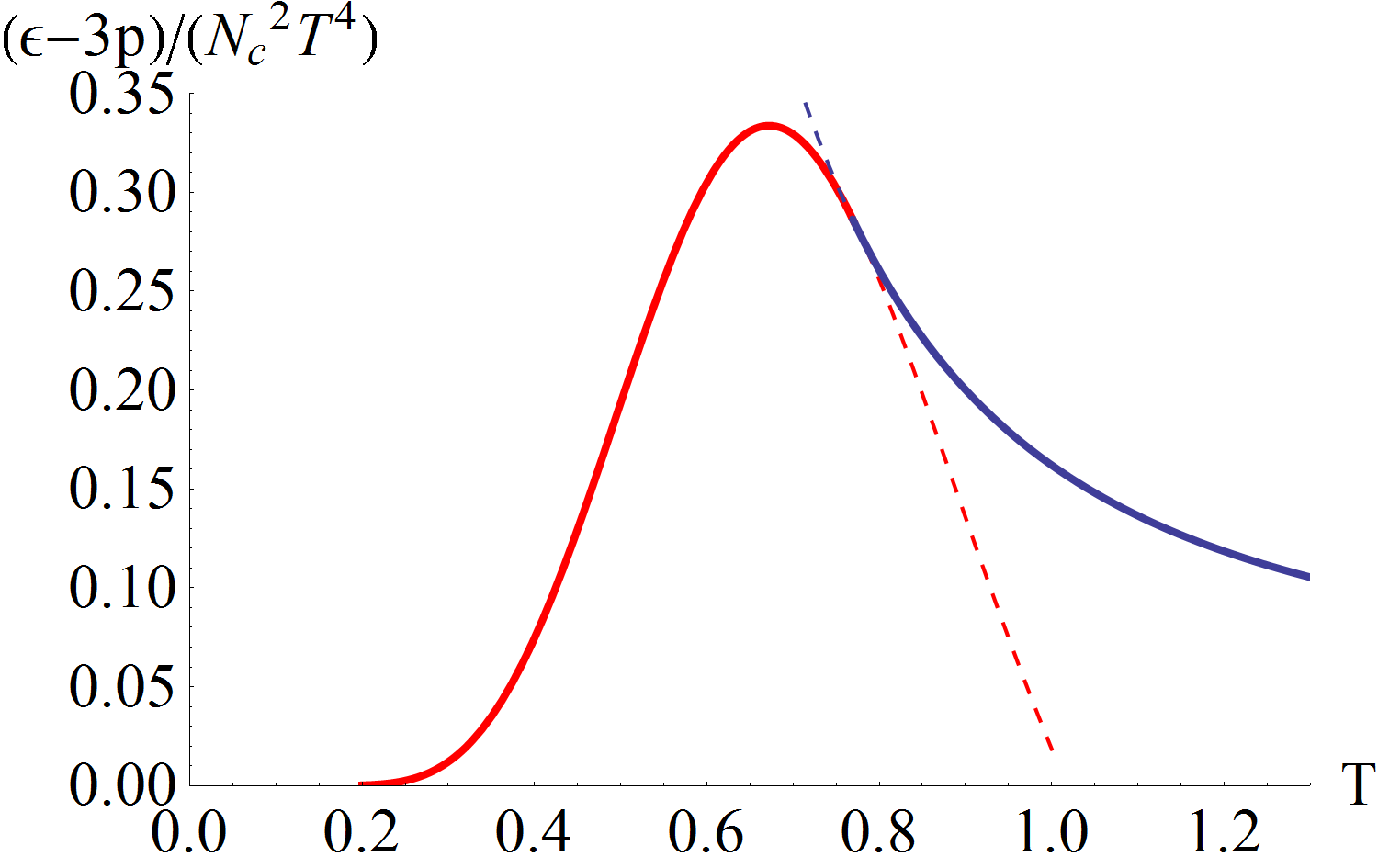}

\end{center}

\caption{\small Pressure and interaction measure with a 3rd order phase transition at $x_f=1$, i.e.,
with $N_f^2=N_c^2$ massless Goldstone bosons.
Stable phases are continuous, metastable ones dashed. The maximum
of interaction measure is at $T=0.672$ in the hadron gas phase,
hadron gas is the stable phase for $T<0.771$ and ends at $T=1.0$. }
\la{fin1}
\end{figure}

\begin{figure}[!t]
\begin{center}

\includegraphics[width=0.49\textwidth]{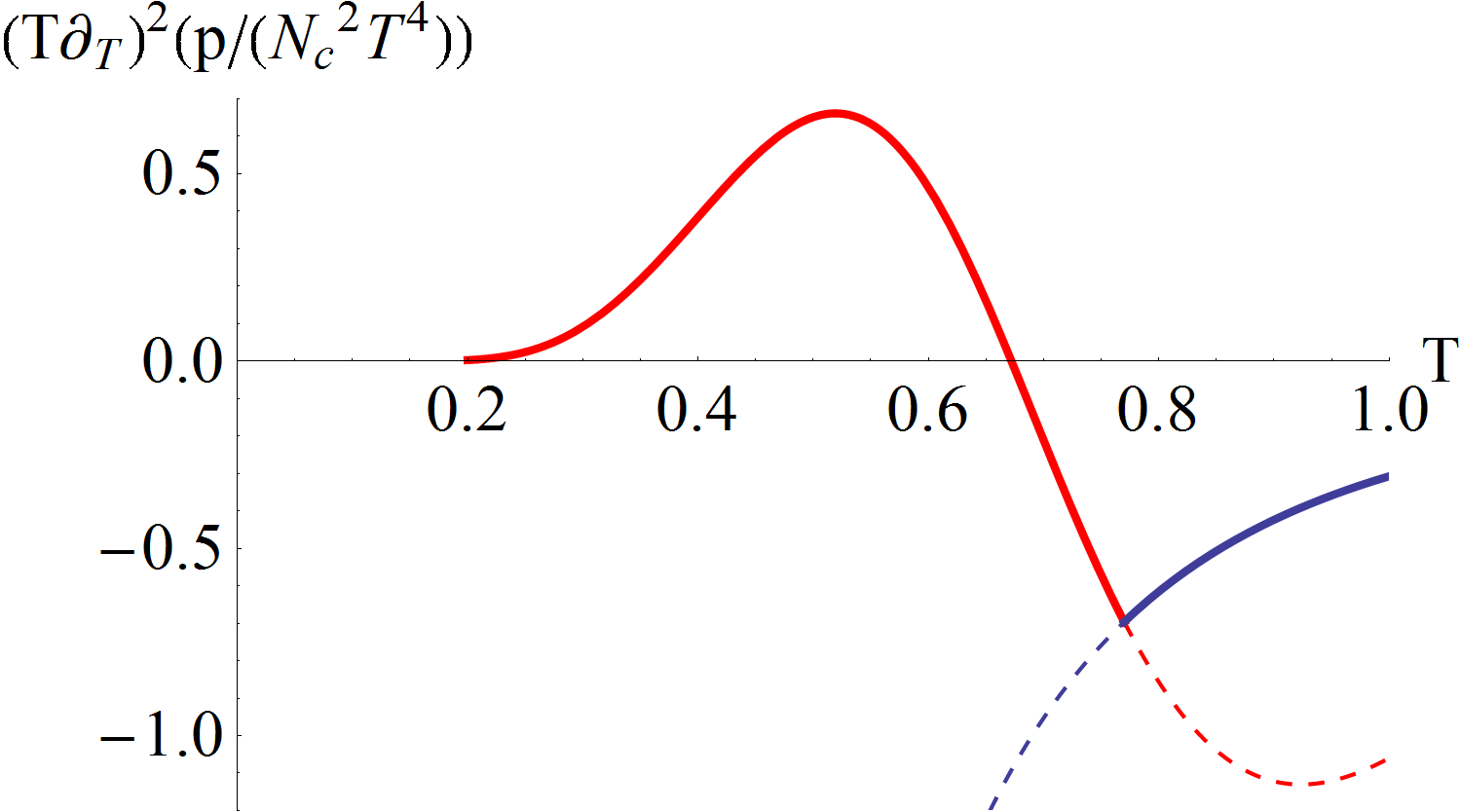}\hfill
\includegraphics[width=0.49\textwidth]{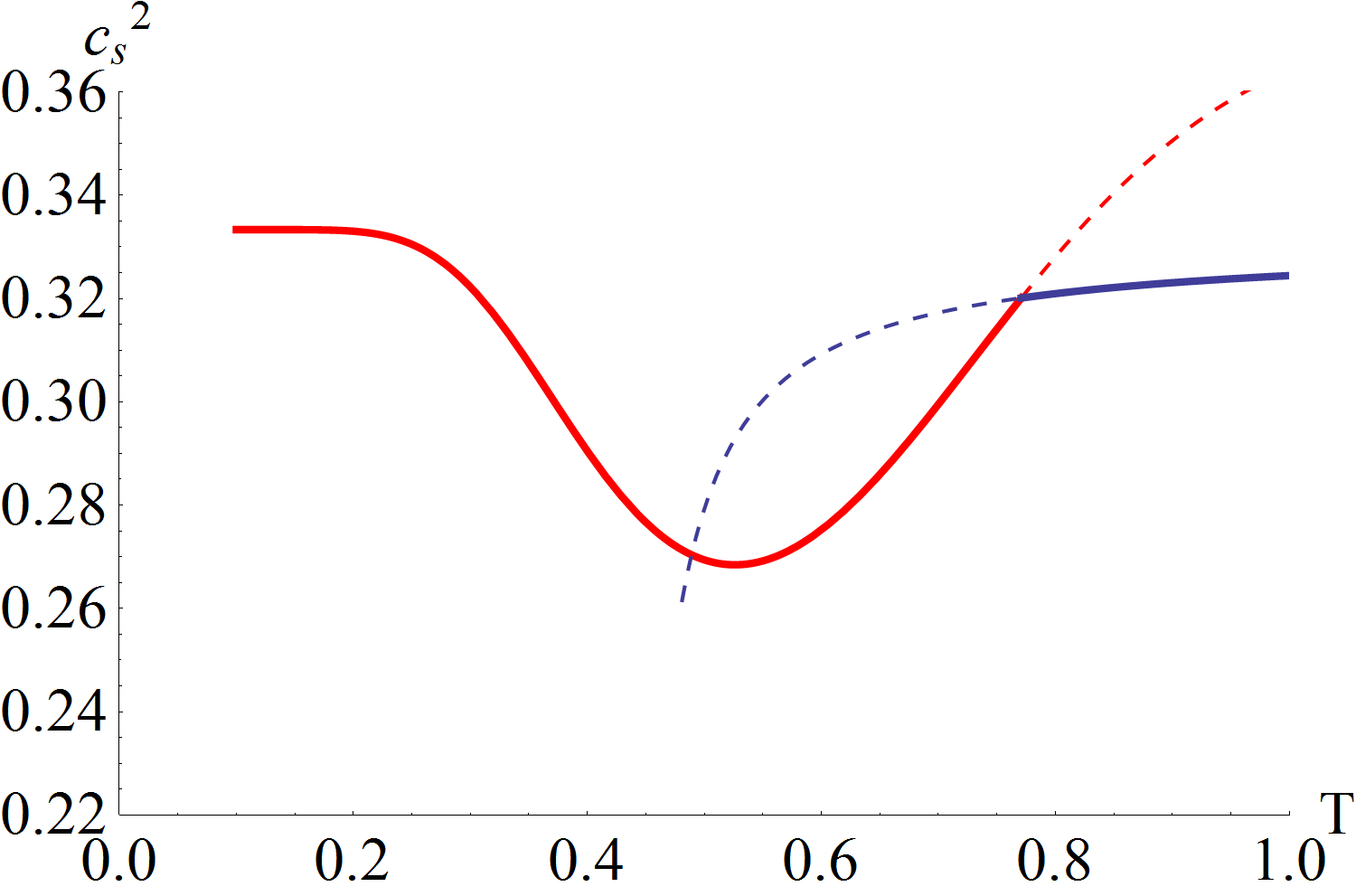}

\end{center}

\caption{\small Second $\log T$ derivative of $p/T^4$ and the sound velocity squared
with a 3rd order phase transition at $x_f=1$. One sees concretely how the 2nd derivative
is continuous at $T_c=0.771$ but the
third derivative jumps. The sound velocity squared is continuous at $T_c$ but not its
derivative. }
\la{fin2}
\end{figure}

\section{Fit of parameters and their $x_f$ dependence}
\la{sect:fin}
We shall now analyse the QCD equation of state (EoS) at $x_f=1,\,2,\,2.5$
by requiring continuity of the logarithmic derivatives $(T\p/\p T)^n(p/T^4)$, with $n=0,1,2$.
We use $p_q$ computed from
holography and the hadronic phase EoS
\be
\ph(T,b,a,\rho_0,m_0)\equiv {p_\rmi{h}\over N_c^2T^4}={\pi^2\over 90}x_f^2
+{\rho_0\over m_0}x_f^2\int_{m_0}^\infty dm\,{m^a\over m_0^a} \,
e^{bm}\,{T_0^3\over T^3}\,W\biggl({m^2T\over 2\pi^2T_0^3}K_2\left({m\over T}\right)\biggr)\,,
\la{phadex}
\ee
where $W(a)$ is as discussed above in \nr{qa}. The first and second $\log T$ derivatives
can be computed analytically but lead to lengthy expressions.

Of particular interest is to see what happens at larger $x_f$, for values approaching
the lower limit of the conformal region. The naive argument comparing the number 
of degrees of freedom in the conformal limits
at $T=0$ and $T=\infty$ gives the estimate $x_c=4$. The precise value
for the potentials of the bulk action used here can be obtained numerically as 
explained in~\cite{jk} and is $x_c=3.187$. The value $2.5$ is already
rather close to this, but not yet in the Miransky scaling region $x_c-x_f \ll 1$, 
where the bound state masses and critical temperatures are expected to decrease as 
$\sim\exp(-{\rm const}/\sqrt{x_c-x_f})$.

The outcome of a numerical application of the excluded volume model for $x_f=1$ is shown in
Figs.~\ref{fin1} and \ref{fin2}. One plots $\hat p$, its first and second derivatives and
the sound velocity, related to the second derivative as in Eq.~\nr{h2nd}.
Thus indeed a connection between hadron
gas in the plasma phase with a third order transition can be established. Not surprisingly, for
a wide range of $T$, $0.7\lsim T\lsim 0.9$, the pressures of the hadron and plasma phases
are very close to each other. The role of the repulsive interactions in the hadron gas
phase was to bend down the 2nd derivative to negative values. It vanishes when the
interaction measure has a maximum. The sound velocity approaches the conformal limit
$1/\sqrt3$ both when $T\to0$ and $T\to\infty$. With broken chiral symmetry and massive
Goldstone bosons, $c_s\to0$ at $T\to0$.

\begin{figure}[!t]
\begin{center}

\includegraphics[width=0.49\textwidth]{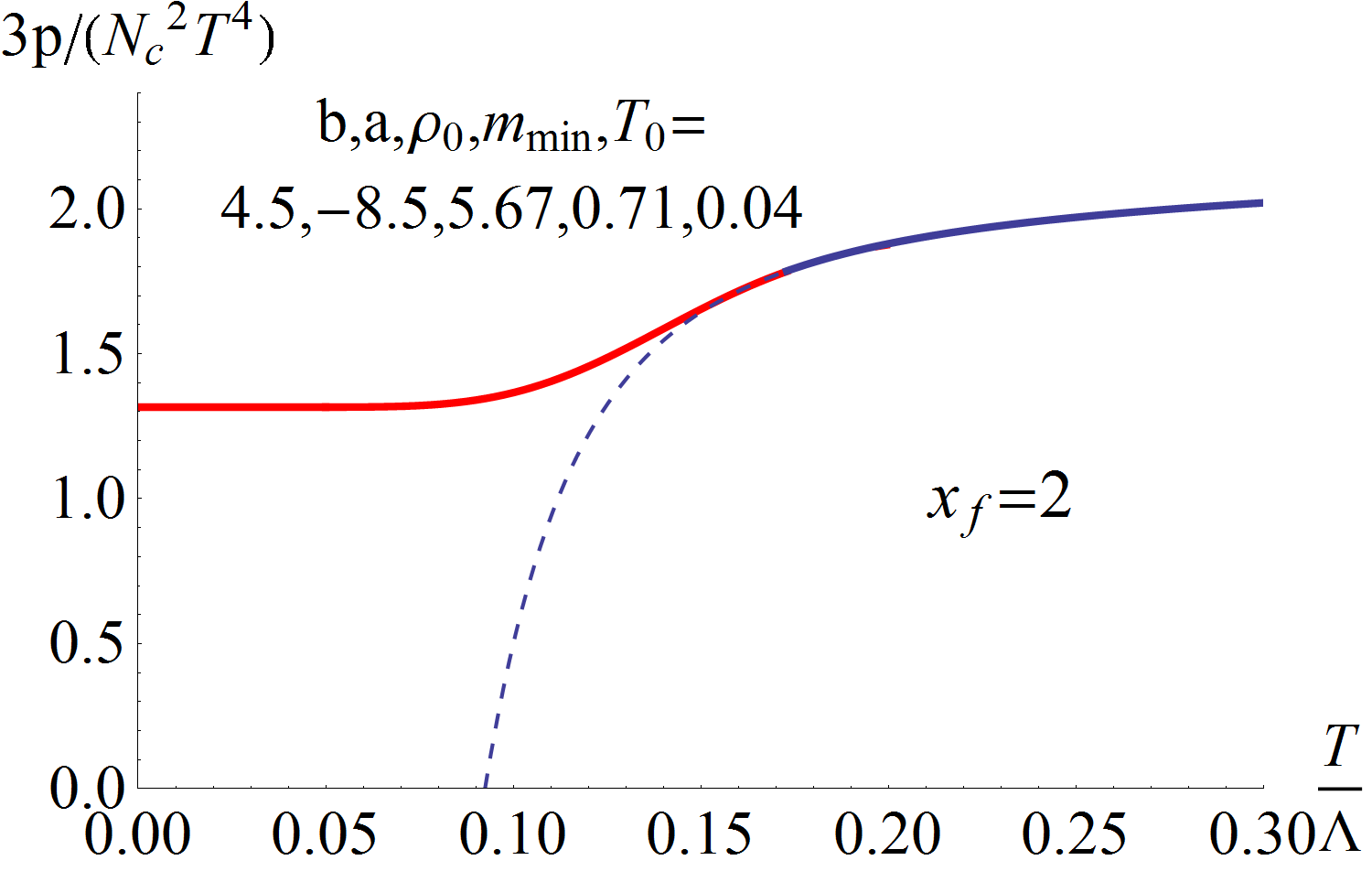}\hfill
\includegraphics[width=0.49\textwidth]{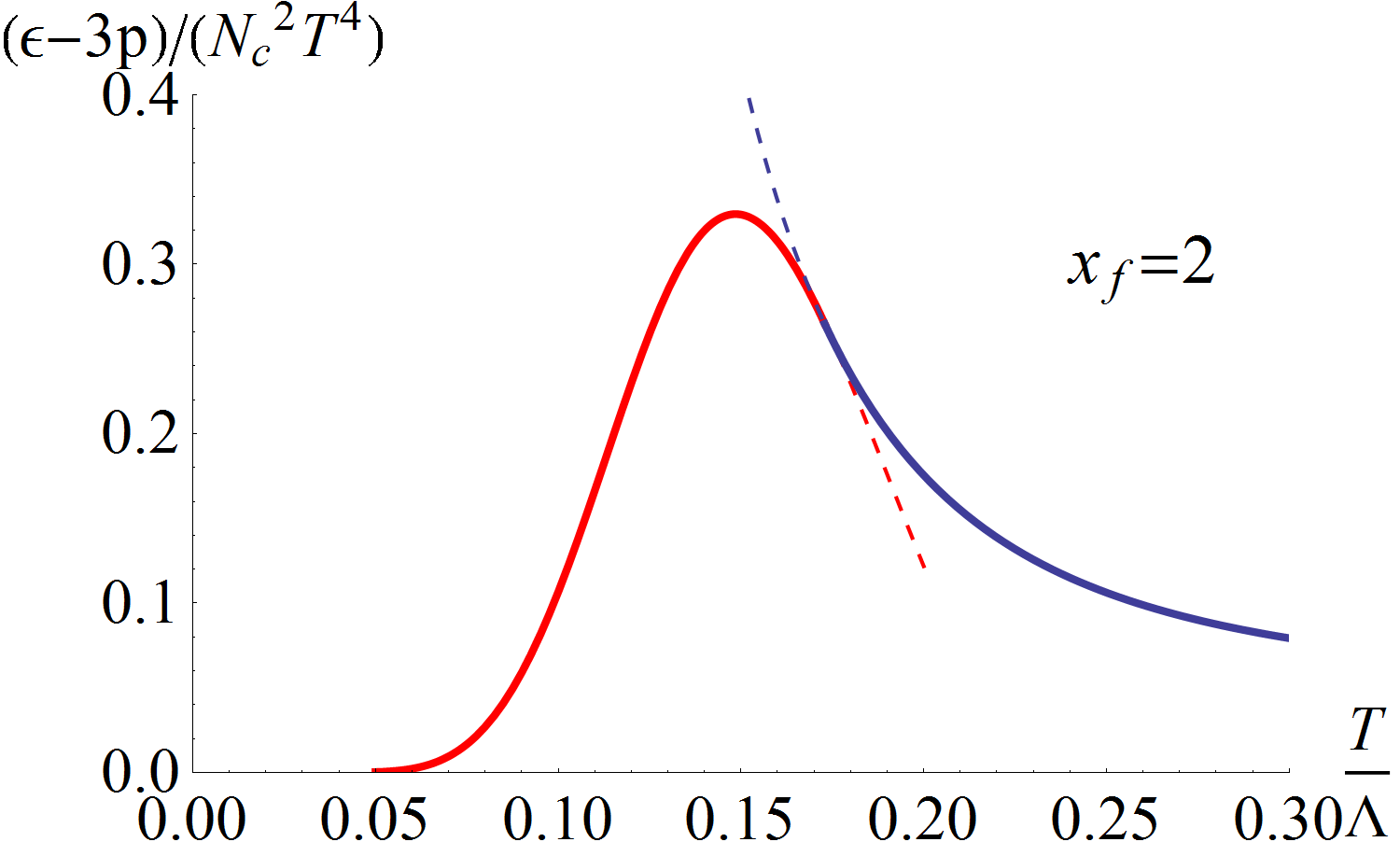}

\end{center}

\caption{\small As Fig.~\protect\ref{fin1} but for $x_f=2$.
Hadron gas ends at $T=0.22$, is the stable phase for $T<0.173$, the maximum
of interaction measure is at $T=0.147$. Note that if we fix the critical temperature to 
the QCD value $T_c\approx 0.15$~GeV, we have here $\Lambda \approx 1$~GeV, i.e., 
the numerical values for the temperatures are close to their physical values measured in units of GeV.}
\la{fin12}
\end{figure}

\begin{figure}[!tb]
\begin{center}

\includegraphics[width=0.49\textwidth]{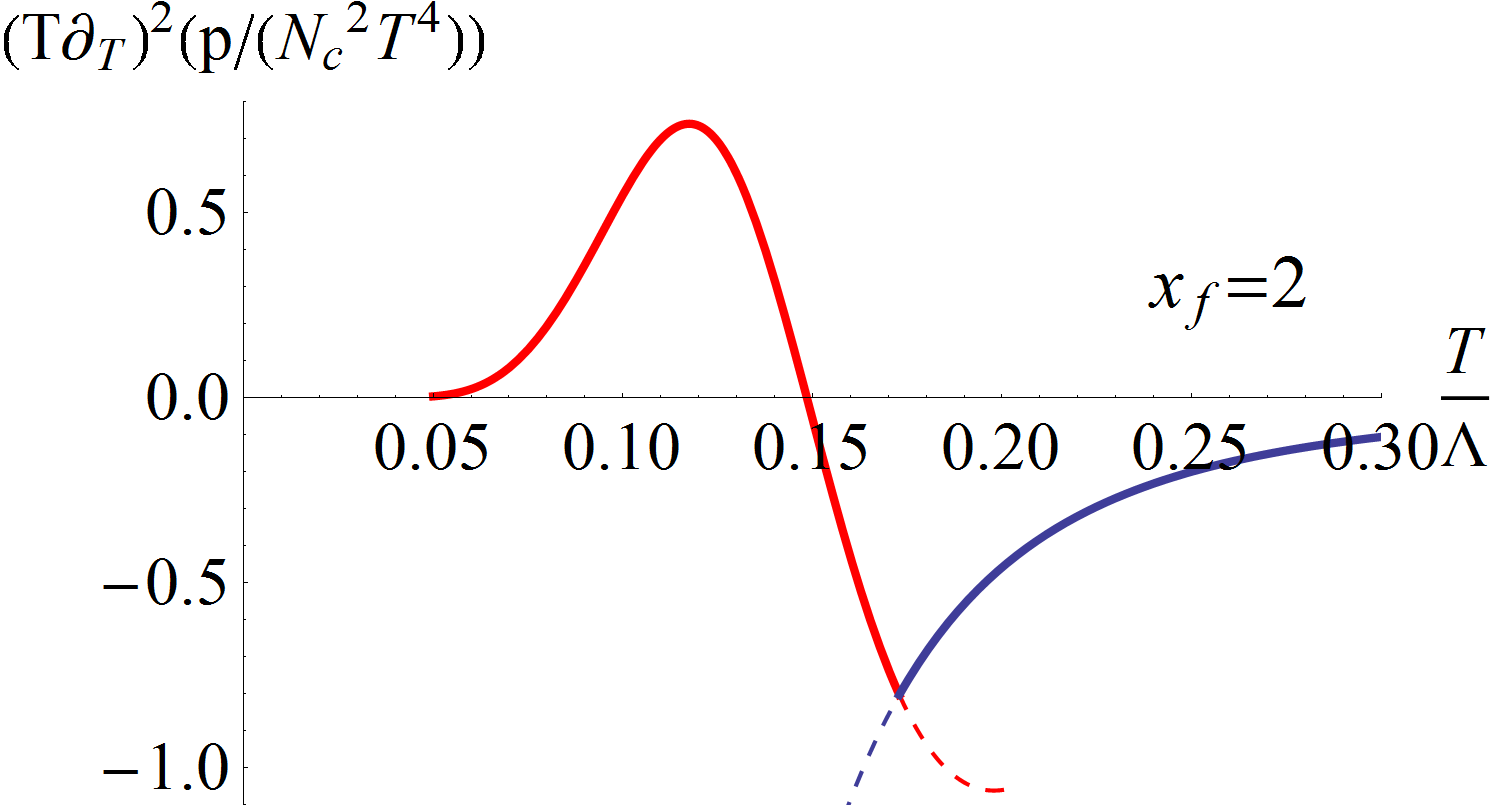}\hfill
\includegraphics[width=0.49\textwidth]{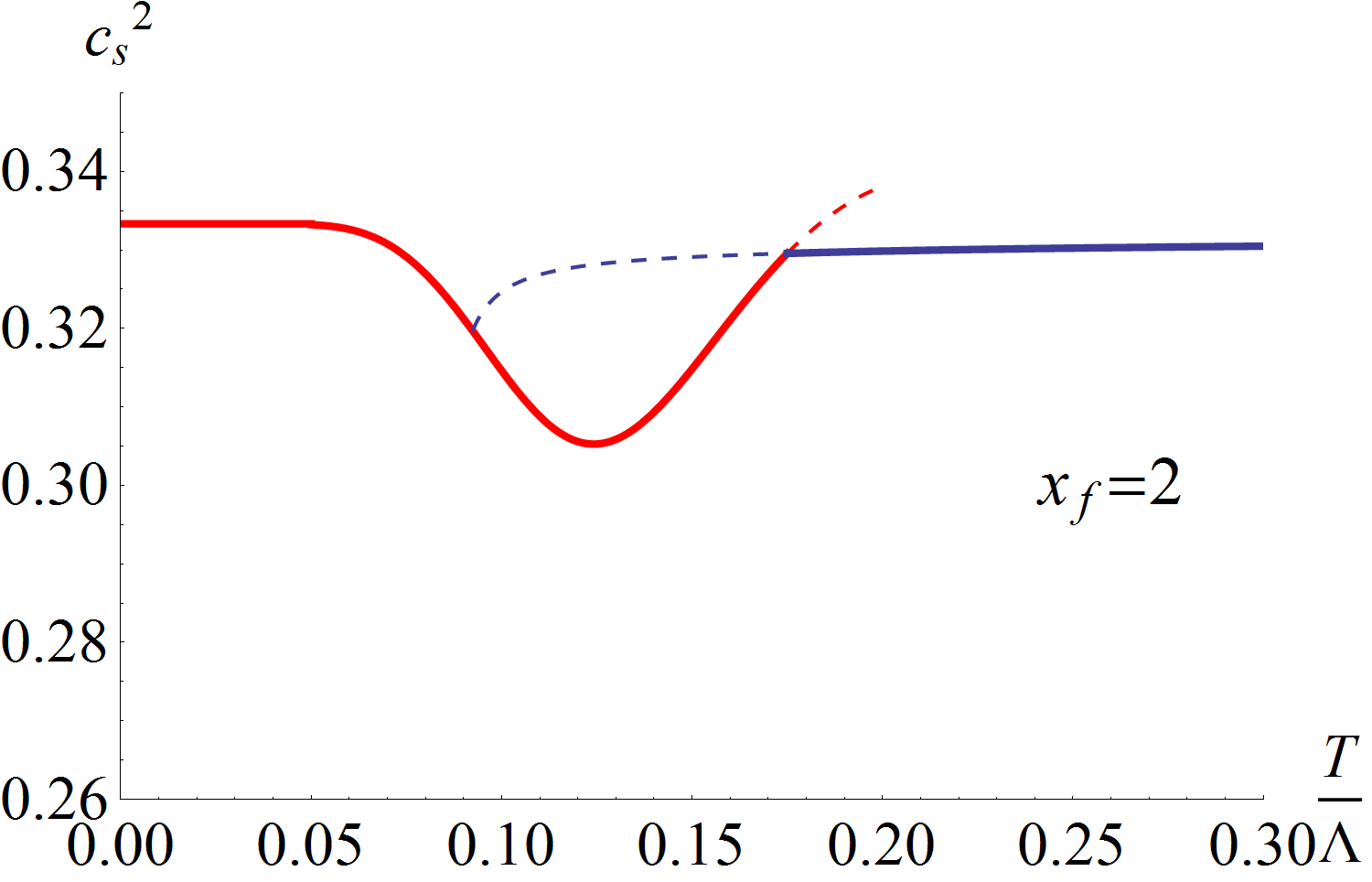}

\end{center}

\caption{\small As Fig.~\protect\ref{fin2} but for $x_f=2$. The 2nd derivative
is continuous at $T_c=0.173$ but the
third derivative jumps. The sound velocity squared is continuous at $T_c$ but its
derivative has a discontinuity. }
\la{fin22}
\end{figure}

The outcome  at $x_f=2$ is shown in Figs.~\ref{fin12} and \ref{fin22} and that
for $x_f=2.5$ in Fig.~\ref{fin22p5} (where only $p$ and the sound velocity are
plotted, 1st and 2nd derivatives are qualitatively as in Figs.~\ref{fin12} and \ref{fin22}).
The fit parameters with mass dimension
are summarised in Table~\ref{tab1}. Here $m_0\equiv m_\rmi{min}$ is the smallest
vector mass, $T(p_q=0)$ is the temperature at which the plasma pressure
vanishes (this is the transition temperature for pure Yang-Mills theory),
$T(p_q=p_\rmi{gb})$ is the temperature at which the plasma pressure is
$p_\rmi{gb}=N_f^2\pi^2T^4/90$, at $T_\rmi{max}$ the interaction measure has
a maximum, at $T_c$ chiral symmetry is restored (the plasma phase becomes
the stable phase), $T_0$ is the meson volume parameter
($v_0\equiv 1/T_0^3$) and $1/b$ is the
Hagedorn temperature (the metastable hadron gas phase ends there).

Although $m_0$ varies by more than a factor $35$, 
the expectation that parameters with mass dimension
scale with $m_0$  is born out to a 10\% accuracy. An exception
is the meson size parameter $T_0$ which decreases by more than expected by mass scaling.
Comparing with $T_c$ we have $T_c/T_0= 3.1,\,4.3,\,4.6$ for $x_f=1,\,2,\,2.5$, respectively.
One may note that even at $x_f=1$ the value of $T_0=0.25$
was smaller that the expected value $0.42$. Thus with increasing $x_f$ the mesons
have to appear effectively larger, the interactions stronger, to bend the hadron gas
EoS nearly continuously to the plasma one.

The numerical value of the mass exponent in $\rho(m)\sim e^{bm}$ can be understood
by noting that physically the hadron gas as a metastable phase cannot be expected
to extend far into the plasma phase. The end point is the Hagedorn temperature
$T_H=1/b$ so that $1/b\gsim T_c\approx 0.3m_0$, in agreement with the observation
$bm_0\approx3$.

The dimensionless parameters vary on a similar level: ($a,\,\rho_0$) =
($-9.61,6.25$) for $x_f=1$ is changed to ($-8.5,\,5.67$) for $x_f=2$ and to ($-8.1,\,6.27$)
for $x_f=2.5$. It is not excluded that constant $x_f$ independent values could be
found for these. Anyway it seems that the exponent of the powerlike mass dependence
is stably close to $-8$ and the weight of the massive part of the mass spectrum is
about $6$.

Note the small range of variation in $c_s^2$, which gets monotonically smaller when $x_f$
increases: one is approaching the
conformal region where everywhere $c_s=1/\sqrt3$.

\begin{figure}[!tb]
\begin{center}

\includegraphics[width=0.49\textwidth]{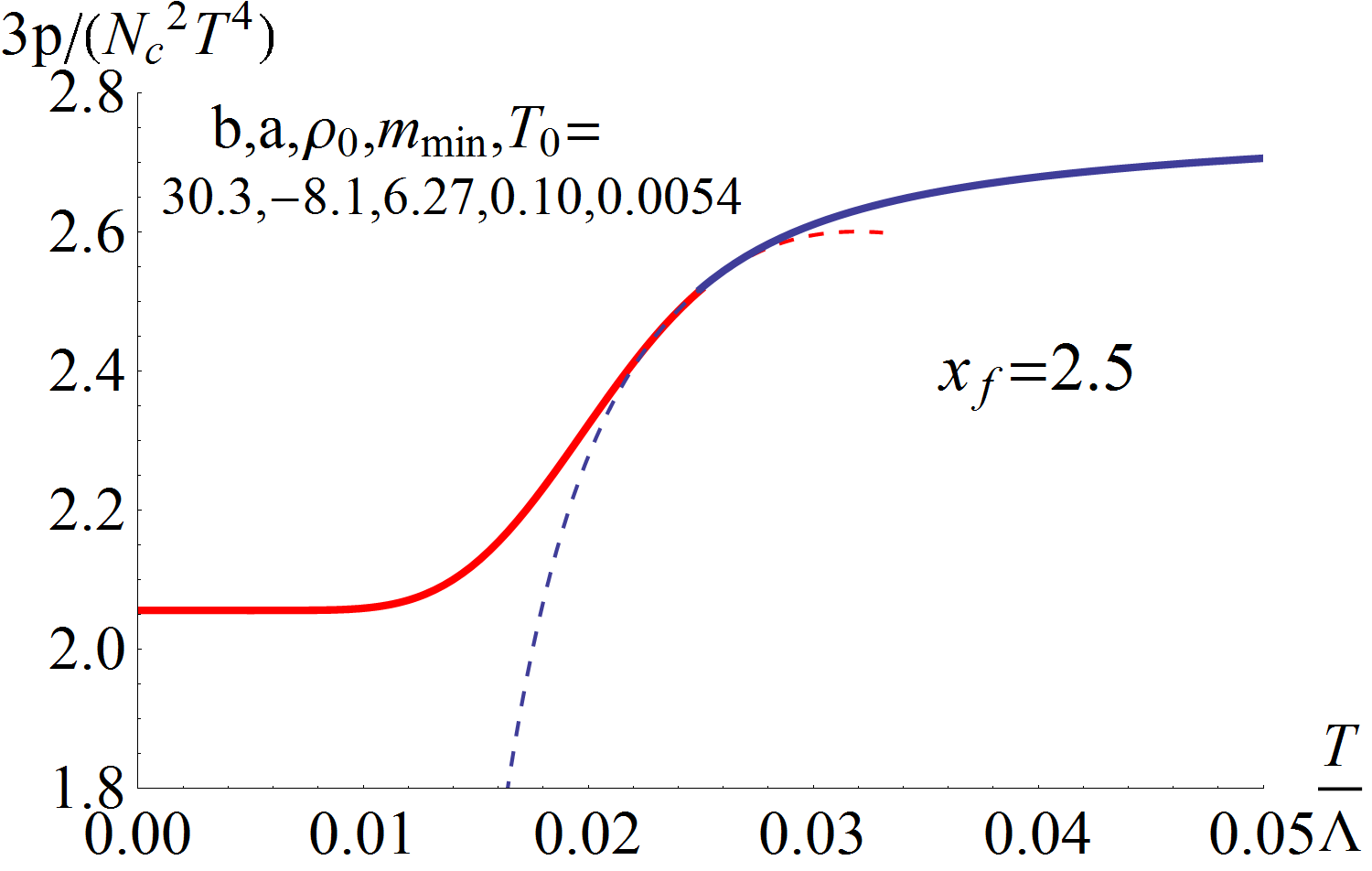}\hfill
\includegraphics[width=0.49\textwidth]{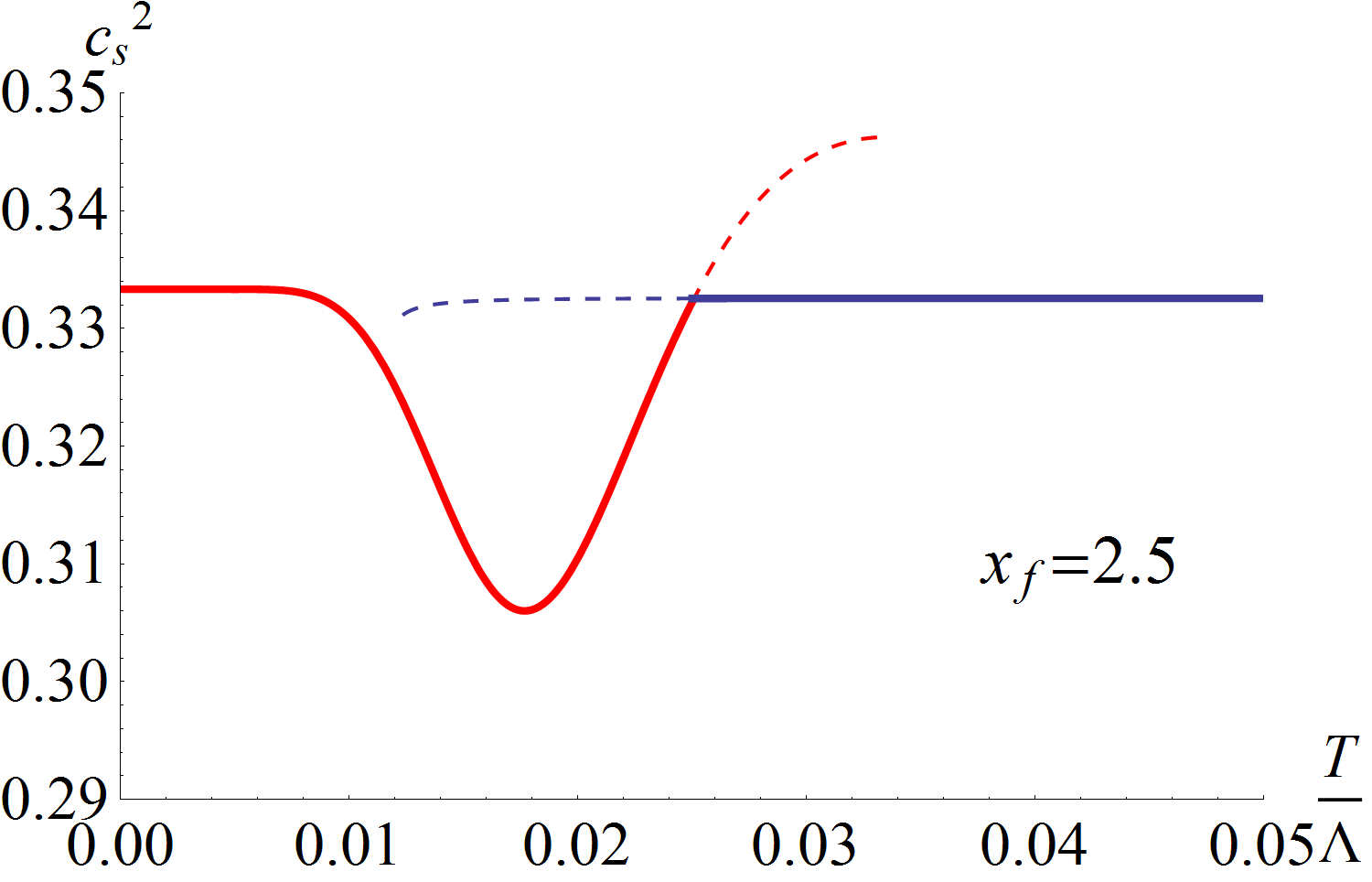}

\end{center}

\caption{\small Left: Fit with a 3rd order phase transition for $x_f=2.5$: the 2nd derivative of
$p$ is continuous at $T_c=0.025$ but the
third derivative jumps. The maximum of the interaction measure (not shown) is a $T_\rmi{max}=0.0209$.
Right: The sound velocity squared. It is continuous at $T_c$ but not its
derivative. }
\la{fin22p5}
\end{figure}

\begin{table}
\begin{center}
\begin{tabular}{|c|c|c|c|c|c|c|c|}
\hline
$x_f$ &$m_0$ &  $T(p_q=0)$ &  $T(p_q=p_\rmi{gb})$ & $T_\rmi{max}$ & $T_c$ &  $T_0$ & $b$\\
\hline
1  &2.82 & 0.480 & 0.542 & 0.672 & 0.771 &  $0.25$ &1 \\
\hline
2  & 0.707 &  0.0925 & 0.125 & 0.147 & 0.173 &  $0.04$  & 4.5                               \\
\hline
2.5 & 0.0795 &  0.0124 & 0.0179 & 0.0209 & 0.025 &  $0.0054$ & 30.3\\
\hline
\end{tabular}

\vspace{5mm}
\begin{tabular}{|c|c|c|c|c|c|c|c|}
\hline
$x_f$ &$m_0$ &  $\hat T(p_q=0)$ &  $\hat T(p_q=p_\rmi{gb})$ & $\hat T_\rmi{max}$
& $\hat T_c$ &  $\hat T_0$ & $bm_0$\\
\hline
1  &2.82 & 0.170 & 0.194 & 0.238 & 0.273 &  $0.0887$ &2.8 \\
\hline
2  & 0.707 &  0.131 & 0.177 & 0.208 & 0.244 &  $0.0566$  & 3.2                               \\
\hline
2.5 & 0.0795 &  0.156 & 0.225 & 0.263 & 0.314 &  $0.0679$ & 3.0\\
\hline
\end{tabular}

\end{center}
\vspace{-5mm}
\caption{Top: Dependence of mass scales on $x_f$. Here $m_0$ is the minimum vector mass.
All quantities are in units of $\Lambda$.
Bottom: Dependence of mass scales on $x_f$ if scaled with $m_0$: $\hat T\equiv T/m_0$. }
\la{tab1}
\end{table}

One may convert the above temperatures to GeV units by, for example, demanding that
$T_c(x_f=1)=0.15$~GeV and by assuming that $\Lambda$ is independent of $x_f$. 
Then $\Lambda=0.194$~GeV and $T_c=150,\,34,\,4.9$~MeV at
$x_f=1,\,2,\,2.5$, respectively. For the determination of scales at large $N_f$ 
and estimates of the $N_f$ dependence of the critical temperature by using other methods,
see, for example, \cite{Jarvinen:2010ks,Braun:2010qs,lombardo}.

\section{Second order transition with massive Goldstone bosons}\la{sect:mq}
If quark masses are non-zero, they break explicitly chiral symmetry
and Goldstone bosons are massive. The hadronic pressure will then vanish at $T=0$.
To see what this implies quantitatively, take the hadron mass spectrum in
\nr{mspect}, use the toy model parameters in \nr{modpar} and replace $\delta(m)\to\delta(m-m_\rmi{gb})$
with $m_\rmi{gb}=0.5\approx T_c$. The argument here is that for physical QCD $T_c$ is close to $m_\pi$.
Also the holographic plasma part will, in principle, change; there are only tachyonic chiral
symmetry breaking solutions even though the quark mass is very small.
We do not yet have these solutions available and stick to the $m_q=0$ plasma curves.
The thermodynamics computed with these assumptions is shown in Fig.~\ref{massiveGB}.

At small $T$ the effect, of course, is striking. 
One observes further that the minimum mass obtained from thermodynamics is now almost equal to
the one from direct computation in \nr{mspect} and that $T_c$ is only 10\% above
the temperature at which $p=0$. Due to this closeness the derivatives are also
larger and the peak in interaction measure even sharper. 

\begin{figure}[!tb]
\begin{center}

\includegraphics[width=0.49\textwidth]{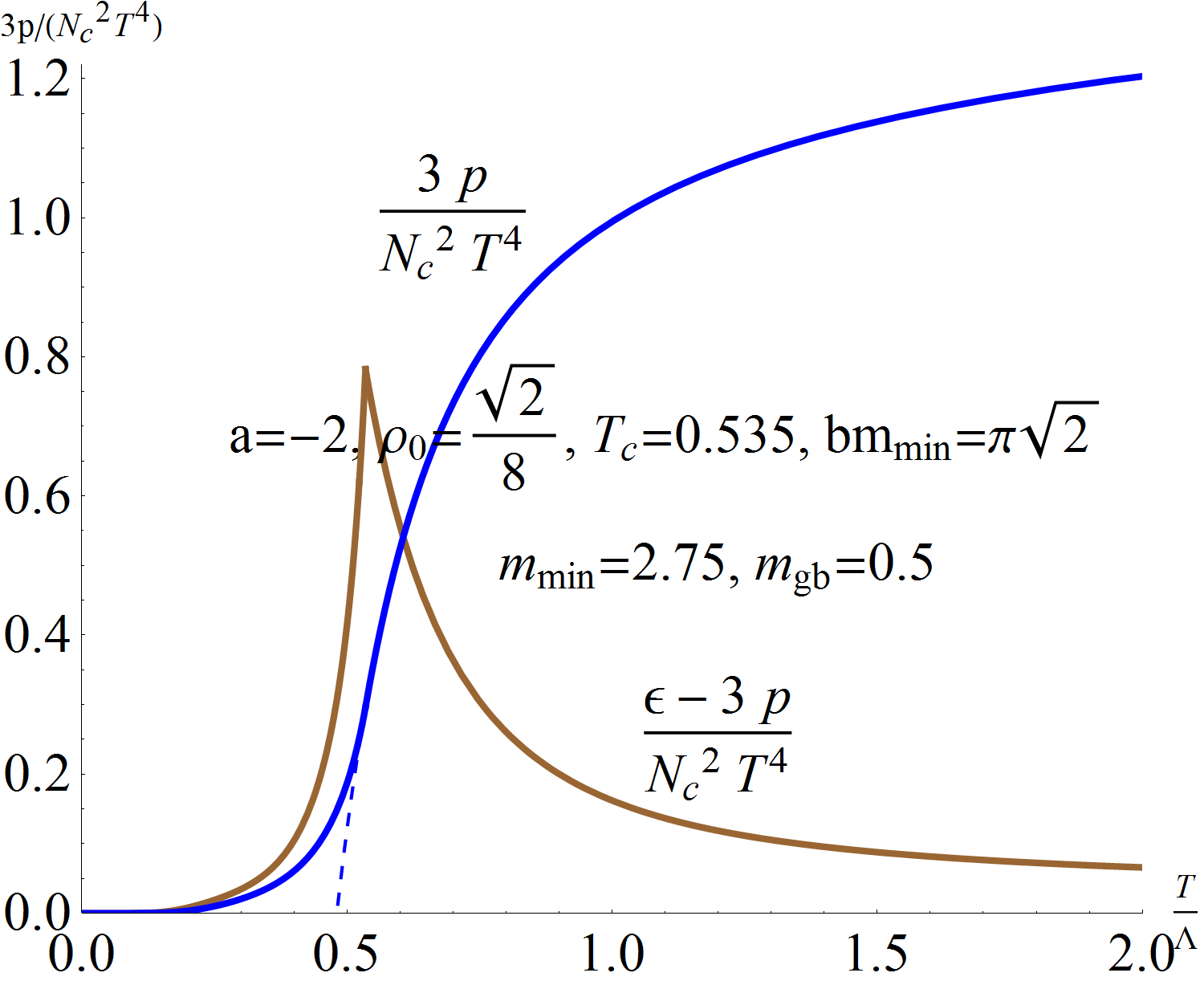}

\end{center}

\caption{\small Left: Thermodynamics with toy model parameters in
\protect\nr{modpar} but with massive Goldstone bosons, $m_\rmi{gb}=0.5\approx T_c$.
The fitted values are $T_c=0.5348$, $m_0=2.754$. 
}
\la{massiveGB}
\end{figure}

\section{Conclusions}
We have in this paper shown concretely how a high $T$, $\mu=0$, QCD plasma equation of state, 
computed from holography at vanishing quark mass, can be connected with a low $T$ hadron gas phase 
with $N_f^2$ Goldstone bosons and massive mesons obeying a Hagedorn spectrum with a minimum mass. 
In the language of holography, the leading holographic computation has been improved
by quantum 1loop and stringy corrections. The holographic computation implies that there is a
minimum temperature for the plasma phase and, accordingly, a phase transition is needed.
It is very simple to connect the phases with a first or second order transition, and we have
shown how a more continuous third order transition can be achieved.
The motivation for this is that for physical non-zero quark masses 
lattice Monte Carlo results suggest that the transition is continuous.

Quantitatively a determining role in the hadron gas phase is played by the minimum mass of
the hadrons (mesons in our case). We emphasize that here these minimum masses have also been 
computed from the same holographic model at $T=0$ and at various values of $x_f=N_f/N_c$.
The entire spectrum obviously cannot be computed.
We have seen here how it and its interactions are constrained 
by the requirement of as continuous a transition as possible.

An outcome of the computations here is that the thermal parameters with mass dimension scale with the
minimum mass when $x_f$ is varied. This is expected to also be true in the Miransky
scaling region, i.e., very close to the start of the conformal region at $x_f=x_c\approx4$.

Our computation, of course, does not prove that the transition is of third order, it just
indicates what phenomena are encountered if this is the case. Conceivably one could
also impose continuity of the third derivative. Completely analytic expression
\cite{kapusta2014} would, however, require that the plasma EoS extend to $T=0$ and the
hadron gas one to $T=\infty$, which is not possible in our model.

Ultimately, of course, these theoretical ideas should be tested by numerical lattice Monte
Carlo simulations, say, at $N_c=3$ and $N_f=3,\,6,\,9,..$ as approximations to $N_c,\,N_f\to\infty$.
Much work has been devoted to this at $T=0$. The figures above should give a good idea of what
can be expected to happen to thermodynamics when $x_f$ is increased. 
The overriding difficulty is the imposition of vanishing or small quark mass. 
This holographic computation can also be extended to non-zero $\mu$.

\vspace{0.5cm}
{\it Acknowledgements}. We thank J. Kapusta for discussions.
This work was supported in part by European Union's Seventh Framework Programme
under grant agreements (FP7-REGPOT-2012-2013-1) No 316165,
PIF-GA-2011-300984, the ERC Advanced Grant BSMOXFORD 228169, and the EU program
Thales MIS 375734. It was also co-financed by the European Union (European Social
Fund, ESF) and Greek national funds through the Operational Program ``Education
and Lifelong Learning'' of the National Strategic Reference Framework (NSRF)
under ``Funding of proposals that have received a positive evaluation in the 3rd and
4th Call of ERC Grant Schemes'', as well as under the action ``ARISTEIA''.
The research of T.A. is supported in part by Icelandic Research Fund grant
130131-052 and by a grant from the University of Iceland Research Fund.


\end{document}